\newif\ifmnrasclass
\IfFileExists{openjournal.cls}{%
  \documentclass[twocolumn]{openjournal}%
}{%
  \IfFileExists{mnras.cls}{\documentclass[fleqn,usenatbib]{mnras}\mnrasclasstrue}%
                          {\documentclass[twocolumn]{article}\usepackage[authoryear,round]{natbib}}%
}
\usepackage{amsmath,amssymb,bm}
\usepackage{graphicx}
\usepackage{hyperref}

\providecommand{\affiliation}[1]{\thanks{#1}}
\ifmnrasclass\else
  \providecommand{\keywords}[1]{\noindent\textbf{Keywords:} #1}
\fi

\newcommand{\vk}{\mathbf{k}}
\newcommand{\vx}{\mathbf{x}}
\newcommand{\intk}{\int_{\vk}}
\newcommand{\nv}{\bm{\nu}}
\newcommand{\avg}[1]{\left\langle #1 \right\rangle}
\newcommand{\dd}{\mathrm{d}}
\newcommand{\iu}{\mathrm{i}}            % imaginary unit
\newcommand{\Snoise}{S^h}              % power spectrum of the reconvolved noise
\newcommand{\Sd}{S^{d}}                 % deconvolved noise spectrum S/|p|^2
\newcommand{\Ssym}{S^{h}_{\rm sym}}      % spin-2-free reconvolved spectrum
\newcommand{\Nadd}{N'}                 % added-noise power spectrum (deconvolved plane)
\newcommand{\Rop}{\mathsf{R}_{90}}
\newcommand{\Kopb}{\mathcal{D}}
\newcommand{\Kop}{\Kopb_1}           % pixel shear-response operator
\newcommand{\Mmat}{\mathsf{M}}
\newcommand{\metacal}{\textsc{Metacalibration}}
\newcommand{\metadet}{\textsc{Metadetection}}
\newcommand{\BFD}{\textsc{BFD}}
\newcommand{\anacal}{\textsc{AnaCal}}

\begin{document}

\title{Low-variance renoising for weak gravitational lensing shear estimation}

\newcommand{\abstracttext}{%
The ``renoising'' noise-bias correction used by \metacal{} and \anacal{} adds to every galaxy image an independent noise field with the statistics of the original noise rotated by $90^\circ$. The rotated field cancels, to first order in shear, the anisotropy that the shear-response operation induces in the original noise, at the cost of doubling the noise variance. We derive the general condition that an added noise field must satisfy for this cancellation: the distribution of the total noise---the original noise transformed with the pixel shear-response operator, plus the added noise carrying an \emph{assigned} shear dependence that we are free to choose---must be independent of shear to first order. Under this condition the shear response of the renoised ellipticity is estimated from the known realization of the added noise alone. The rotated noise field, which we call rotation renoising, is one solution of the condition and yields the analytical response estimator of \anacal{}. We then construct a second solution, spectral renoising, which adds only a small fraction of the original noise variance: for homogeneous Gaussian noise the added field is an independent, scaled-down copy of the image noise, and its assigned shear response is the same field passed through a fixed filter set by the noise power spectrum deconvolved by the point spread function. Image simulations show that spectral renoising keeps the multiplicative shear bias at the $10^{-3}$ level and raises the effective galaxy number density by about 20 per cent at the faint end.
}
\newcommand{\keywordstext}{gravitational lensing: weak --- techniques: image processing --- methods: statistical}

\ifmnrasclass
  % mnras: author + affiliation in one argument; abstract and keywords
  % (an environment) come after \maketitle.
  \author[X. Li]{Xiangchong Li\\
    Brookhaven National Laboratory, Bldg 510, Upton, NY 11973, USA}
  \maketitle
  \begin{abstract}
  \abstracttext
  \end{abstract}
\else
  \author{Xiangchong Li}
  \email{xli6@bnl.gov}
  \affiliation{Brookhaven National Laboratory, Bldg 510, Upton, NY 11973, USA}
  % \author{...}
  % \affiliation{...}
  \begin{abstract}
  \abstracttext
  \end{abstract}
  \maketitle
\fi

\section{Introduction}
\label{sec:intro}

Large-scale photometric imaging surveys are opening a new window on precision cosmology. A principal science goal of these programs is to test the standard cosmological model and to elucidate the physical origin of cosmic acceleration. Weak gravitational lensing by the large-scale structure of the Universe---the coherent distortion that intervening matter imprints on the observed shapes of distant galaxies---offers one of the most direct probes of the matter distribution and its growth over time \citep{bartelmannSchneider2001, kilbinger2015, bartelmannMaturi2017, mandelbaum2018, pratBacon2025}. Over the past decade, the ``Stage III'' surveys have demonstrated the cosmological power of this probe through tomographic cosmic-shear analyses, including the Dark Energy Survey \citep{abbott2022, abbott2026}, the Kilo-Degree Survey \citep{liKiDS2023, wright2025} and the Hyper Suprime-Cam (HSC) Subaru Strategic Program \citep{liHSC2023, dalal2023}.

Three ``Stage IV'' programs will together image billions of galaxies over the next decade: the Vera C.\ Rubin Observatory Legacy Survey of Space and Time \citep[LSST,][]{ivezic2019}, the European Space Agency's \textit{Euclid} mission \citep{laureijs2011}, and NASA's Nancy Grace Roman Space Telescope \citep{spergel2015}. Their science requirements impose tight bounds on the multiplicative shear bias, typically at the $\sim\!10^{-3}$ level \citep{massey2013, lsstdesc2018}, motivating the development of ever more rigorous calibration methods.

The lensing distortion is a remapping of image coordinates: light that would have arrived from the source-plane position $\bm\beta$ in the absence of lensing is observed at the image-plane position $\bm\theta$. In the weak-lensing regime the deflection varies slowly across a galaxy image, and to leading order the mapping is locally linear,
\begin{equation}
  \bm\beta = \Mmat\,\bm\theta,
  \qquad
  \Mmat = \begin{pmatrix}
    1-\kappa-\gamma_1 & -\gamma_2 \\
    -\gamma_2 & 1-\kappa+\gamma_1
  \end{pmatrix},
  \label{eq:jacobian}
\end{equation}
where $\kappa$ is the convergence, which magnifies or demagnifies the image isotropically, and $(\gamma_1,\gamma_2)$ are the two components of the shear, which stretch the image anisotropically: $\gamma_1$ along the $x_1$ and $x_2$ axes, $\gamma_2$ along the diagonals. A galaxy with surface brightness $f(\bm\beta)$ in the source plane is observed as $f_\gamma(\bm\theta)=f(\Mmat\bm\theta)$ before convolution with the point spread function (PSF). In this paper we assume $\kappa=0$ and focus on the shear components. More precisely, the quantity that galaxy shapes can constrain is the reduced shear $g\equiv\gamma/(1-\kappa)$, which reduces to $\gamma$ in the limit $\kappa\to0$ adopted here. In practice the shear is extracted statistically from the observed ellipticities of an ensemble of many galaxies, under the assumption that their intrinsic shapes are randomly oriented, so that the ensemble average of the intrinsic ellipticity vanishes and is not correlated with the shear direction, and an accurate estimate requires the linear shear response of the shape measurement, i.e. the first-order derivative of the measured shape with respect to the applied shear \citep{huffMandelbaum2017}.

Several approaches now calibrate this response without relying on external image simulations, at least when blended sources lie at the same redshift. \metadet{} \citep{sheldonHuff2017, sheldon2020} achieves this through numerical self-calibration of detection and selection effects, while the Bayesian Fourier Domain method \citep[\BFD{};][]{bernsteinArmstrong2014, bernstein2016} frames shear inference as a Bayesian problem over the Fourier moments of the galaxy light, though it may still require percent-level simulation input to correct detection biases in blended fields. The \anacal{} framework \citep{liMandelbaum2023, liMandelbaum2024} computes the response analytically, by propagating the analytical shear response of each pixel of the PSF-deconvolved, Gaussian-smoothed image through detection and measurement with the chain rule; it reaches accuracy at second order in shear, and its analytical nature makes it exceptionally efficient, processing each detected galaxy in under a millisecond.

Since the lensing distortion acts on the intrinsic galaxy light before the image noise, the noise does not carry the shear. The pixel shear response, however, is computed from the noisy galaxy image, with operations on the image as a whole, which treats the noise as if it carried the shear. Quantities that are linear in the pixel values are not affected by this: the noise averages to zero, so their expectation is that of the noiseless image and their shear response is unchanged. Galaxy ellipticities are nonlinear in the pixel values, and for them the noise does not average out: the expectation of a nonlinear estimator over the noise differs from its noiseless value, and this ``noise bias'' depends on the shear and biases the shear estimate \citep{kaiser2000, refregier2012}. One way to remove it is to correct the estimator analytically, deriving the noiseless expectations of both the shape and its shear response from the noisy estimates by an expansion in the noise \citep{refregier2012, liMandelbaum2023}. The other way, followed by the \metacal{} framework \citep{huffMandelbaum2017, sheldonHuff2017} is to estimate the shear response directly from the noisy data: the image is deconvolved, sheared, reconvolved, and the response of the estimator is measured by finite differences. Because the shear operation is applied to the noise as well as to the galaxy, the noise in the sheared images acquires an anisotropy. \citet{sheldonHuff2017} remove it by adding to the image an independent noise field with the statistics of the original noise rotated by $90^\circ$ (``renoising''). \citet{liMandelbaum2024} (hereafter LM24) gave an analytical derivation of the renoised shear response within the \anacal{} framework \citep{liMandelbaum2023}, where the response is obtained analytically from the shear response of the smoothed pixel values rather than by finite differences. In both frameworks the added field doubles the image noise, which reduces the effective galaxy number density by 20--25 per cent (LM24). Deep-field variants of both frameworks---deep-field \metacal{} \citep{zhang2023} and deep-field \anacal{} \citep{park2025}---have been proposed to recover this loss by measuring the shear response on deep-field images, whose much lower noise makes the added field correspondingly weaker, at the cost of requiring deep-field data. This paper is inspired by \citet{sheldon2026}, who showed that \metacal{} can reach an unbiased response by adding only the small noise deficit that symmetrizes the shear-induced modulation of the noise power; here we develop the analogous construction for the analytical response of \anacal{}.

We observe that the added noise only needs to cancel the first-order change of the noise statistics under the shear operation. We therefore treat the shear dependence of the added field as a quantity we are free to assign, and derive the condition under which the resulting response estimator is unbiased. The rotated noise field satisfies it, and so does a much weaker noise field: an independent copy of the image noise scaled down by an adjustable factor, whose assigned shear response is the same field passed through a fixed filter built from the shear derivative of the logarithm of the PSF-deconvolved noise power spectrum. We call this \emph{spectral renoising}. It adds only a small fraction of the original noise variance to every galaxy image; the price is a noisier shear response, which is acceptable because the response enters the shear estimate only through its average over the galaxy sample. We validate both forms of renoising on LSST-like image simulations with realistic blending: the multiplicative bias stays within the LSST ten-year requirement $|m|<2\times10^{-3}$ \citep{lsstdesc2018, sheldon2023} in every configuration tested, and spectral renoising raises the effective galaxy number density by about $20$ per cent at the faint end relative to rotation renoising.

Section~\ref{sec:general} states the condition and the resulting response estimator. Section~\ref{sec:rotation} shows that the rotated noise field, which we call \emph{rotation renoising}, satisfies the condition and recovers the LM24 estimator. Section~\ref{sec:spectral} constructs the spectral-renoising field, whose variance is an adjustable small fraction of the original noise variance, shows that it satisfies the condition, and shows how to remove the spin-2 part of the noise spectrum induced by an anisotropic PSF with little added noise before the construction is applied. Section~\ref{sec:tests} tests both forms of renoising on image simulations. Section~\ref{sec:discussion} summarizes the results and sketches directions for future work. Throughout, we follow the notation of LM24.

\section{Shear response of renoised images}
\label{sec:general}

This section recalls the smoothed image and the pixel shear response of \anacal{} (Section~\ref{sec:observables}), shows that an ellipticity response measured from a noisy image acquires a spurious shear dependence through the noise (Sections~\ref{sec:ellip} and~\ref{sec:noisy}), and states the condition on an added noise field that cancels it, together with the resulting response estimator (Section~\ref{sec:condition}).

\subsection{Smoothed image and pixel shear response}
\label{sec:observables}

We follow the notation of \citet{li2025}. Let $f^p(\vx)$ be the observed image, i.e.\ the galaxy convolved with the PSF plus image noise, and $p(\vx)$ the PSF model, with $\vx=(x_1,x_2)$ the image-plane (pixel) coordinate. A field and its Fourier transform are written with the same symbol and distinguished by the argument,
\begin{equation}
  f^p(\vk) = \iint\dd^2x\; f^p(\vx)\,\exp(-\iu\,\vk\cdot\vx),
  \label{eq:ft}
\end{equation}
where $\iu$ is the imaginary unit ($\iu^2=-1$), so that $\partial f^p(\vx)/\partial x_j$ transforms to $\iu k_j f^p(\vk)$; $\mathcal{F}^{-1}$ denotes the inverse transform, and $\intk\equiv\iint\dd^2k/(2\pi)^2$ the corresponding Fourier-space integral. \anacal{} does not work with $f^p$ directly but with the \emph{smoothed image}, obtained by deconvolving the PSF and reconvolving with a round Gaussian \citep{liMandelbaum2023},
\begin{equation}
  f^h(\vk) = h(\vk)\,\frac{f^p(\vk)}{p(\vk)},
  \qquad
  h(\vk) = \exp\!\left(-\frac{k^2\sigma_h^2}{2}\right),
  \label{eq:smoothed}
\end{equation}
where $k\equiv|\vk|$, and the width $\sigma_h$ is chosen so that the Gaussian is broader than the PSF, which keeps the deconvolution well conditioned. The lensing shear acts on the galaxy before the PSF, so under the remapping (\ref{eq:jacobian}) with $\kappa=0$ the deconvolved image transforms as
\begin{equation}
  \frac{f^p_\gamma(\vk)}{p(\vk)} = \frac{f^p}{p}\!\left(\Mmat^{-\mathsf{T}}\vk\right),
  \label{eq:shear}
\end{equation}
with $\Mmat$ the matrix of equation~(\ref{eq:jacobian}) at $\kappa=0$; this is the Fourier-space form of $f^p_\gamma(\bm\theta)=f^p(\Mmat\bm\theta)$ up to the factor $\det\Mmat=1+\mathcal{O}(\gamma^2)$. Thus $\gamma_1>0$ stretches the image along $x_1$, and $\gamma_2>0$ stretches it along the diagonal $x_1=x_2$. In what follows we set $\gamma_2=0$ and derive the formulae for $\gamma_1$; the corresponding results for $\gamma_2$ are given at the end of this section. The Gaussian $h$ is applied after the shear and is not sheared. To first order the smoothed pixel values therefore change by a fixed linear operator, the analytical pixel shear response of \citet{liMandelbaum2023}. Following \citet[eqs.~5--6]{li2025}, define the auxiliary smoothed images
\begin{align}
  G_1(\vx) &= \sigma_h^2\,\mathcal{F}^{-1}\!\Big[\left(k_1^2-k_2^2\right)f^h(\vk)\Big],
  \nonumber\\
  G_2(\vx) &= \sigma_h^2\,\mathcal{F}^{-1}\!\Big[2k_1k_2\,f^h(\vk)\Big],
  \nonumber\\
  J_j(\vx) &= \iu\,\mathcal{F}^{-1}\!\Big[k_j\,f^h(\vk)\Big] = \frac{\partial f^h}{\partial x_j},
  \qquad j=1,2 ;
  \label{eq:GJ}
\end{align}
the pixel shear response is then
\begin{equation}
  (\Kop f^h)(\vx) \equiv \frac{\partial f^h(\vx)}{\partial\gamma_1} = G_1(\vx) - x_1 J_1(\vx) + x_2 J_2(\vx),
  \label{eq:pixres}
\end{equation}
which defines the operator $\Kop$. $G_1$ characterizes how the Gaussian kernel responds to a shear applied between deconvolution and reconvolution, and the $J$ terms describe how the shear displaces neighbouring pixels relative to one another (Appendix~\ref{app:shear}). In Fourier space $(\Kop f^h)(\vk)=(k_1\partial_{k_1}-k_2\partial_{k_2})f^h(\vk)+\sigma_h^2(k_1^2-k_2^2)f^h(\vk)$. $\Kop$ applies to any smoothed field, and it is a spin-2 operator: with $\Rop$ the rotation of the coordinates by $90^\circ$, $\Rop\Kop\Rop^{-1}=-\Kop$. The second shear component gives, by the same steps with $\gamma_1=0$ in equation~(\ref{eq:shear}), the operator $\Kopb_2$,
\begin{equation}
  (\Kopb_2 f^h)(\vx) \equiv \frac{\partial f^h(\vx)}{\partial\gamma_2} = G_2(\vx) - x_1 J_2(\vx) - x_2 J_1(\vx),
  \label{eq:pixres2}
\end{equation}
which in Fourier space reads $(\Kopb_2 f^h)(\vk)=(k_1\partial_{k_2}+k_2\partial_{k_1})f^h(\vk)+2\sigma_h^2k_1k_2\,f^h(\vk)$; it is $\Kop$ in the frame rotated by $45^\circ$. Throughout, the formulae are written for $\gamma_1$ and $\Kop$; the corresponding results for $\gamma_2$ follow by replacing $\Kop$ with $\Kopb_2$, and the final estimator for $\gamma_2$ is given in Section~\ref{sec:measured}.

\subsection{Shear response of the ellipticity}
\label{sec:ellip}

A galaxy ellipticity $e_1$ is a nonlinear function of the smoothed pixel values, $e_1=e_1(f^h)$, whether it is built from moments \citep{li2018,liLiMassey2022}, from model fitting \citep{li2025}, from a convolutional neural network with $D_4$ symmetry \citep{lin2026, lin2026b} or otherwise. Write $e_{1,\vx}\equiv\partial e_1/\partial f^h(\vx)$ for its derivative with respect to the value of the pixel at $\vx$. Its shear response follows from the chain rule over pixels,
\begin{equation}
  R_1 \equiv \frac{\partial e_1}{\partial\gamma_1}
  = \sum_{\vx} e_{1,\vx}\,\frac{\partial f^h(\vx)}{\partial\gamma_1}
  = \sum_{\vx} e_{1,\vx}\,(\Kop f^h)(\vx),
  \label{eq:R1}
\end{equation}
where the sum runs over the pixels; in our implementation the pixel response is propagated through the linear shapelet measurement of the Fourier Power Function Shapelets method \citep[FPFS;][]{li2018,liLiMassey2022} (Appendix~\ref{app:modes}). The shear estimator is $\hat\gamma_1=\avg{e_1}/\avg{R_1}$, and likewise $\hat\gamma_2=\avg{e_2}/\avg{R_2}$ with $R_2=\sum_{\vx}e_{2,\vx}(\Kopb_2 f^h)(\vx)$ and $e_{2,\vx}\equiv\partial e_2/\partial f^h(\vx)$; both are accurate to second order in shear when the measurement has spin-2 symmetry (LM24, Section~2.1).

On a noiseless image equation~(\ref{eq:R1}) is the exact response of the galaxy. On the observed image it is not, because the measurement is made from a noisy image and $\Kop$ is applied to the whole smoothed image, noise included. The observed image is the noiseless galaxy convolved with the PSF plus the image noise, $f^p=\bar f^p+\epsilon$; carried through the same deconvolution and smoothing, the smoothed image splits as $f^h=\bar f^h+\epsilon^h$, with $\bar f^h$ the smoothed noiseless galaxy and $\epsilon^h$ the smoothed image noise (Section~\ref{sec:noisy} gives their statistics). Using the linearity of $\Kop$,
\begin{equation}
  R_1 = \sum_{\vx} e_{1,\vx}\,(\Kop\bar f^h)(\vx) + \sum_{\vx} e_{1,\vx}\,(\Kop\epsilon^h)(\vx).
  \label{eq:R1split}
\end{equation}
Only the first term is wanted. The lensing shear acts on the galaxy light before the image noise is added, so the noise is not sheared; the operator $\Kop$ nevertheless acts on it along with the galaxy, and $\Kop\epsilon^h$ is the response the noise \emph{would} have if it were sheared too. This spurious term does not average away, because the weights $e_{1,\vx}$ are themselves evaluated on the noisy image and are correlated with $\epsilon^h$, and it makes the measured response depend on the shear through the noise---the noise bias of the response. It cannot be subtracted directly, because the realization $\epsilon^h$ is unknown. The rest of the paper removes this spurious shear dependence of the noise by adding a further, controlled noise field, in two ways: rotation renoising (Section~\ref{sec:rotation}) cancels it to first order with a $90^\circ$-rotated independent noise field, and spectral renoising (Section~\ref{sec:spectral}) cancels it in the noise power spectrum with a small, adjustable added variance.

\subsection{Image noise under the pixel shear response}
\label{sec:noisy}

Before constructing the schemes that remove the spurious term $\Kop\epsilon^h$ of equation~(\ref{eq:R1split}), we summarize how the pixel shear response changes the statistical properties of the noise. The relevant quantity is the first-order change that $\Kop$ induces in the noise power spectrum; it is fixed by the PSF and the image-noise spectrum alone, and it is what the added noise must cancel.

The image noise $\epsilon(\vx)$ introduced in Section~\ref{sec:ellip} is added after the PSF and is not convolved with it; we take it to be homogeneous with power spectrum $S(\vk)$. It passes through the same deconvolution and reconvolution as the galaxy, so that the decomposition $f^h=\bar f^h+\epsilon^h$ introduced there reads explicitly
\begin{equation}
  f^h = \bar f^h + \epsilon^h,
  \qquad
  \bar f^h(\vk) = h\,\frac{\bar f^p(\vk)}{p(\vk)},
  \qquad
  \epsilon^h(\vk) = h\,\frac{\epsilon(\vk)}{p(\vk)},
  \label{eq:smoothednoise}
\end{equation}
where $\epsilon^h$ is the \emph{reconvolved noise}: the image noise divided by the PSF in Fourier space, as part of the deconvolution of the image, although it was never convolved with the PSF, and then reconvolved with the Gaussian. This division is why $\epsilon^h$ is not white even for white image noise. It is homogeneous with power spectrum
\begin{align}
  \Snoise(\vk) &= |h(\vk)|^2\,\Sd(\vk),
  \qquad
  \Sd(\vk)\equiv\frac{S(\vk)}{|p(\vk)|^2},
  \nonumber\\
  \avg{\epsilon^h(\vk)\,\epsilon^{h*}(\vk')}&=\Snoise(\vk)\,\delta_{\rm D}(\vk-\vk'),
  \label{eq:Sdec}
\end{align}
where $\Sd$ is the spectrum of the \emph{deconvolved} noise $\epsilon/p$ and $\delta_{\rm D}$ is the Dirac delta function, up to the normalization of the Fourier transform.

The spurious term of equation~(\ref{eq:R1split}) is built from $\Kop\epsilon^h$, the first-order change that the noise \emph{would} undergo if it were transformed together with the galaxy (the smoothed image of a sheared galaxy is $\bar f^h+\gamma_1\Kop\bar f^h+\epsilon^h$, not $(1+\gamma_1\Kop)(\bar f^h+\epsilon^h)$, because the shear acts on the galaxy and not on the noise). The field $\Kop\epsilon^h$ is not observable, because the realization $\epsilon^h$ is unknown, but its statistics are: the field $\epsilon^h+\gamma_1\Kop\epsilon^h$ is homogeneous with power spectrum
\begin{align}
  \Snoise_\gamma(\vk) &= \Snoise(\vk)\big[1+\gamma_1\,a_1(\vk)\big] + \mathcal{O}(\gamma^2),
  \label{eq:Sgamma}\\
  a_1(\vk) &= \left(k_1\frac{\partial}{\partial k_1}-k_2\frac{\partial}{\partial k_2}\right)\ln\Snoise(\vk)
  \nonumber\\
  &\quad+ 2\sigma_h^2\left(k_1^2-k_2^2\right),
  \label{eq:Adef}
\end{align}
where the two terms come from the $J$ and $G_1$ terms of $\Kop$ (Appendix~\ref{app:shear}). The Gaussian factor $|h(\vk)|^2=\exp(-\sigma_h^2k^2)$ gives
\begin{equation}
  \ln\Sd(\vk)=\ln\Snoise(\vk)+\sigma_h^2k^2,
  \label{eq:lnSd}
\end{equation}
so that $(k_1\partial_{k_1}-k_2\partial_{k_2})\ln\Snoise=(k_1\partial_{k_1}-k_2\partial_{k_2})\ln\Sd-2\sigma_h^2(k_1^2-k_2^2)$; the last term cancels the second term of equation~(\ref{eq:Adef}), leaving
\begin{equation}
  a_1(\vk)=\left(k_1\frac{\partial}{\partial k_1}-k_2\frac{\partial}{\partial k_2}\right)\ln\Sd(\vk).
  \label{eq:adef}
\end{equation}
Thus $\gamma_1 a_1(\vk)$ is the fractional change of the noise power at $\vk$ when the noise is transformed with $\Kop$. It is the shear derivative of the logarithm of the deconvolved spectrum $\Sd$ and does not depend on the smoothing Gaussian: because the Gaussian is applied after the shear it contributes no anisotropy of its own, it only limits the band. The function $a_1$ is real and even in $\vk$, and is fixed by the PSF and the image-noise spectrum alone. The second component defines in the same way
\begin{equation}
  a_2(\vk)\equiv\left(k_1\frac{\partial}{\partial k_2}+k_2\frac{\partial}{\partial k_1}\right)\ln\Sd(\vk),
  \label{eq:a2def}
\end{equation}
the shear derivative in the frame rotated by $45^\circ$, so that the spectrum of $\epsilon^h+\gamma_2\Kopb_2\epsilon^h$ is $\Snoise(1+\gamma_2 a_2)$ to first order.

\subsection{The renoising approach and the condition on the added noise}
\label{sec:condition}

The response measured on a noisy image carries the spurious term $\Kop\epsilon^h$ (Section~\ref{sec:ellip}, equation~\ref{eq:R1split}), which cannot be computed because the realization of the image noise $\epsilon^h$ is unknown. The renoising approach makes it computable by adding to the image a second, independent pure-noise field of \emph{known} realization, and assigning that field a shear response that cancels this first-order noise term in the ensemble average; the shear response of the renoised ellipticity is then expressed entirely in computable quantities, and the shear estimator follows. This subsection turns that requirement into a precise condition on the added noise, and Sections~\ref{sec:rotation} and \ref{sec:spectral} construct two fields that satisfy it.

Write $n(\vx)$ for the pure-noise image added to the data before detection and measurement. It passes through the pipeline like everything else and appears in the smoothed image as a field $n^h$, so that the smoothed, renoised image is
\begin{equation}
  f^h = \bar f^h + \epsilon^h + n^h ,
  \label{eq:renoised}
\end{equation}
which is equation~4 of \citet{li2025} with the noise split into its two parts; from here on $f^h$ denotes this renoised image, as there. The pipeline computes the pixel response $\Kop f^h$ of the full renoised image and, from the stored $n$, the response $\Kop n^h$. Since the lensing shear acts on the galaxy and not on either noise field, the shear response of $\avg{e_1}$ is
\begin{align}
  \avg{R_1} &= \avg{\sum_{\vx} e_{1,\vx}\,(\Kop\bar f^h)(\vx)}
  \nonumber\\
  &= \avg{\sum_{\vx} e_{1,\vx}\left(\Kop f^h-\Kop\epsilon^h-\Kop n^h\right)(\vx)},
  \label{eq:Rrenoised}
\end{align}
where the average is over galaxies and both noise fields, and $e_{1,\vx}$ is evaluated on the renoised image. The only unmeasurable term is the one containing $\Kop\epsilon^h$.

To this end we let the added noise transform with the bookkeeping parameter $\gamma_1$ in a way of our own choosing, different from the way the signal transforms: a family of fields $n^h_\gamma$ with $n^h_0=n^h$, built from the same noise realizations for every $\gamma_1$; throughout, a subscript $0$ denotes the member of a family at $\gamma=0$. The family is an auxiliary construction for estimating a derivative, so the assigned dependence need not represent physical lensing: only its member at $\gamma=0$ enters the data, and only its derivative there enters the response. Its \emph{assigned response} is
\begin{equation}
  \dot n^h_{(1)}\equiv\left.\frac{\partial n^h_\gamma}{\partial\gamma_1}\right|_{\gamma=0}.
  \label{eq:assigned}
\end{equation}
Consider the total noise in the smoothed image when the original noise is transformed with the operator $\Kop$ of the signal and the added noise follows its assigned family,
\begin{equation}
  \zeta_\gamma \equiv \epsilon^h + \gamma_1\,\Kop\epsilon^h + n^h_\gamma .
  \label{eq:zeta}
\end{equation}
\textbf{Condition.} \emph{The probability distribution of $\zeta_\gamma$ is independent of $\gamma_1$ to first order.}

Since $\partial\zeta_\gamma/\partial\gamma_1|_0=\Kop\epsilon^h+\dot n^h_{(1)}$ by equation~(\ref{eq:zeta}), the condition implies that this combination, averaged over both noise fields with the measurement weights $e_{1,\vx}$ evaluated on the renoised image at $\gamma=0$, contributes nothing:
\begin{equation}
  \avg{\sum_{\vx} e_{1,\vx}\left(\Kop\epsilon^h+\dot n^h_{(1)}\right)(\vx)}_{\epsilon,n} = 0 .
  \label{eq:identity}
\end{equation}
The first term alone does not vanish---the weights $e_{1,\vx}$ are correlated with the noise, and $\avg{\sum_{\vx} e_{1,\vx}\,\Kop\epsilon^h}$ is the noise bias of the response---but the condition fixes the whole distribution of $\zeta_\gamma$ to first order in $\gamma_1$, so the first-order change of \emph{every} noise statistic, equation~(\ref{eq:identity}) among them, is zero. Inserting it into (\ref{eq:Rrenoised}) gives the response of the renoised ellipticity in computable quantities only:
\begin{equation}
  \avg{R_1} = \avg{\sum_{\vx} e_{1,\vx}\left(\Kop f^h-\Kop n^h+\dot n^h_{(1)}\right)(\vx)}.
  \label{eq:main}
\end{equation}
Here $\Kop n^h$ and $\dot n^h_{(1)}$ are distinct: $\Kop n^h$ is the response the added noise would have if it were sheared with the galaxy, whereas $\dot n^h_{(1)}$ is the response assigned to it in equation~(\ref{eq:assigned}); for the rotated noise field of Section~\ref{sec:rotation} the two are equal and opposite, $\dot n^h_{(1)}=-\Kop n^h$, and in general they are unrelated.

\section{Rotation renoising: the rotated noise field}
\label{sec:rotation}

Rotation renoising cancels the spurious shear dependence of the noise to first order with an independent noise field rotated by $90^\circ$: because the pixel response $\Kop$ is spin-2, the rotated field acquires the opposite response, so that the shear-dependent part of the two noise fields cancels to first order. We first review the method of \citet{sheldonHuff2017} in the analytical form given by LM24 and \citet{li2025}, and then show that it is a solution of the condition of Section~\ref{sec:condition}.

\subsection{Renoising in LM24}
\label{sec:review}

\citet{sheldonHuff2017} take $n=\Rop\epsilon'$, where $\epsilon'$ is an independent realization of the image noise, with the same statistical properties as $\epsilon$ but independent of it, and $\Rop$ rotates by $90^\circ$. The rotated noise is deconvolved with the correspondingly rotated PSF $p^{90}$ and reconvolved with the same Gaussian, so that the smoothed, renoised image is \citep[eq.~4]{li2025}
\begin{align}
  f^h(\vk) &= h(\vk)\left(\frac{f^p(\vk)}{p(\vk)}+\frac{n(\vk)}{p^{90}(\vk)}\right),
  \nonumber\\
  n^h &= h\,\frac{n}{p^{90}} = \Rop\!\left(\frac{h\,\epsilon'}{p}\right) ,
  \label{eq:opt1image}
\end{align}
where the second equality holds because $h$ depends only on $|\vk|$ and $p^{90}$ is $p$ rotated by $90^\circ$, and $h\,\epsilon'/p$ has the same statistics as $\epsilon^h$. LM24 write the expectation of the renoised ellipticity over galaxies and both noise fields and differentiate it with respect to the shear, which gives equation~(\ref{eq:Rrenoised}) (their eqs.~31--35 and~39). To eliminate the unmeasurable term, they argue that, since $\epsilon^h$ and $n^h$ are noise realizations oriented $90^\circ$ relative to each other after deconvolution, and the galaxies are randomly oriented,
\begin{equation}
  \avg{\sum_{\vx} e_{1,\vx}\,(\Kop\epsilon^h)(\vx)}
  = \avg{\sum_{\vx} e_{1,\vx}\,(\Kop n^h)(\vx)}
  \label{eq:lm24eq40}
\end{equation}
(their eq.~40), which turns equation~(\ref{eq:Rrenoised}) into the renoised response estimator (their eq.~41)
\begin{equation}
  \avg{R_1} = \avg{\sum_{\vx} e_{1,\vx}\,\Big(\Kop\big[f^h-2n^h\big]\Big)(\vx)}.
  \label{eq:lm24}
\end{equation}
The field $\Kop[f^h-2n^h]=\Kop[h(f^p/p-n/p^{90})]$ is exactly the combination of $G_1$, $J_1$ and $J_2$ built from $f^p/p-n/p^{90}$ in equation~6 of \citet{li2025}. The price is that the added field carries the rotated power $\Snoise(\Rop\vk)$, with the same integrated pixel variance as the original noise, so the noise variance doubles.

\subsection{Cancelling the first-order shear dependence}
\label{sec:opt1solution}

In the language of Section~\ref{sec:condition}, assign to the rotated noise field the family
\begin{equation}
  n^h_\gamma = n^h - \gamma_1\,\Kop n^h,
  \qquad
  \dot n^h_{(1)} = -\,\Kop n^h,
  \label{eq:opt1family}
\end{equation}
with $\Kop$ the pixel shear-response operator of equation~(\ref{eq:pixres}). Because $\Kop$ is a spin-2 operator, $\Rop\Kop\Rop^{-1}=-\Kop$, so $n^h_\gamma=\Rop(1+\gamma_1\Kop)(h\,\epsilon'/p)$: the total noise $\zeta_\gamma$ is the sum of the original noise transformed with $\Kop$ and an independent, $90^\circ$-rotated field with the same statistics as that transformed noise. For homogeneous noise whose probability law is invariant under $90^\circ$ rotation---for Gaussian noise this is the invariance of the spectrum, $\Snoise(\Rop\vk)=\Snoise(\vk)$, which fixes the whole law---the condition of Section~\ref{sec:condition} holds for any such noise distribution, Gaussian or not: writing $\zeta_\gamma=(1+\gamma_1\Kop)\,u+\Rop(1+\gamma_1\Kop)\,v$ with $u=\epsilon^h$ and $v=h\,\epsilon'/p$ independent and identically distributed, the exchange $(u,v)\to(\Rop v,\,\Rop^{-1}u)$ preserves their joint law and, because $\Kop\Rop=-\Rop\Kop$, maps $\zeta_\gamma$ to $\zeta_{-\gamma}$; the distribution of $\zeta_\gamma$ is therefore even in $\gamma_1$, and its first-order change vanishes. For homogeneous noise without $90^\circ$ symmetry the covariance of $\zeta_\gamma$ is not invariant, but the identity (\ref{eq:identity}) still holds after averaging over an isotropic galaxy population; this is the content of Appendix~B of LM24.

With $\dot n^h_{(1)}=-\Kop n^h$ the identity (\ref{eq:identity}) is exactly equation~(\ref{eq:lm24eq40}), and substituting into equation~(\ref{eq:main}) recovers the estimator (\ref{eq:lm24}).

\subsection{No spin-2 part in noise after rotation renoising}
\label{sec:opt1spin2}

Section~\ref{sec:opt1solution} removed the shear dependence of the noise to first order, so the estimator is unbiased to first order. Its second-order accuracy rests on one further property of the noise at $\gamma=0$. The ellipticity is a spin-2 quantity, so by the spin argument of LM24 (their eqs.~16--18) it gains a noise-induced additive bias only if the \emph{unsheared} noise is itself anisotropic in a spin-2 way---that is, if its power differs between a mode $\vk$ and the perpendicular mode $\Rop\vk$.

At $\gamma=0$ the renoised noise is $\zeta_0=\epsilon^h+n^h$, the original noise plus the independent $90^\circ$-rotated field, with covariance
\begin{equation}
  \Snoise(\vk)+\Snoise(\Rop\vk),
  \label{eq:opt1sym}
\end{equation}
the sum of the two spectra because the fields are independent. This is invariant under $\vk\to\Rop\vk$ for \emph{any} image-noise spectrum and any PSF, since $\Rop^2\vk=-\vk$ and the power spectrum of a real field is even, $\Snoise(-\vk)=\Snoise(\vk)$, so the unsheared noise has no spin-2 part in its covariance. For Gaussian noise the covariance fixes the distribution, so this symmetry of the spectrum is the symmetry of the full law that the spin argument requires; for non-Gaussian noise the covariance alone does not constrain the higher-order statistics, and the distributional symmetry holds when the law of the image noise is itself invariant under $90^\circ$ rotation, for then $\zeta_0=u+\Rop v$ with $u$, $v$ independent and identically distributed is manifestly symmetric. Adding the full rotated field therefore symmetrizes the noise, and the measured spin-2 ellipticity carries no additive bias correlated with the anisotropy of the noise or of the PSF; this is why rotation renoising is accurate to second order for an arbitrary PSF. The price is that the full rotated field doubles the noise variance; Section~\ref{sec:spin2} shows how spectral renoising reaches the same symmetry at a much smaller cost.

\section{Spectral renoising: low-variance added noise}
\label{sec:spectral}

Rotation renoising cancels the shear dependence of the noise in configuration space, by a rotation, and pays for it with a full extra noise field of the same variance. Spectral renoising instead cancels it mode by mode in the noise power spectrum, which lets the added variance be made as small as a fraction $\mu$ of the original, with $\mu>0$ a free parameter that we keep small. We now assume homogeneous Gaussian noise with reconvolved spectrum $\Snoise(\vk)$, which need not have any symmetry. The construction below rests on this assumption; in Section~\ref{sec:tests}, however, we also test the accuracy of the method under source photon noise---realized as Gaussian noise whose variance tracks the galaxy light, so that the test probes the inhomogeneity of the noise---in a setup at the ten-year LSST depth (Y10). Sections~\ref{sec:construction}--\ref{sec:measured} construct the low-variance added noise for a spectrum that is invariant under $90^\circ$ rotations, and Section~\ref{sec:spin2} extends it to a general spectrum by first removing its spin-2 part.

\subsection{Construction of the added noise}
\label{sec:construction}

\begin{figure*}
  \centering
  \includegraphics[width=\textwidth]{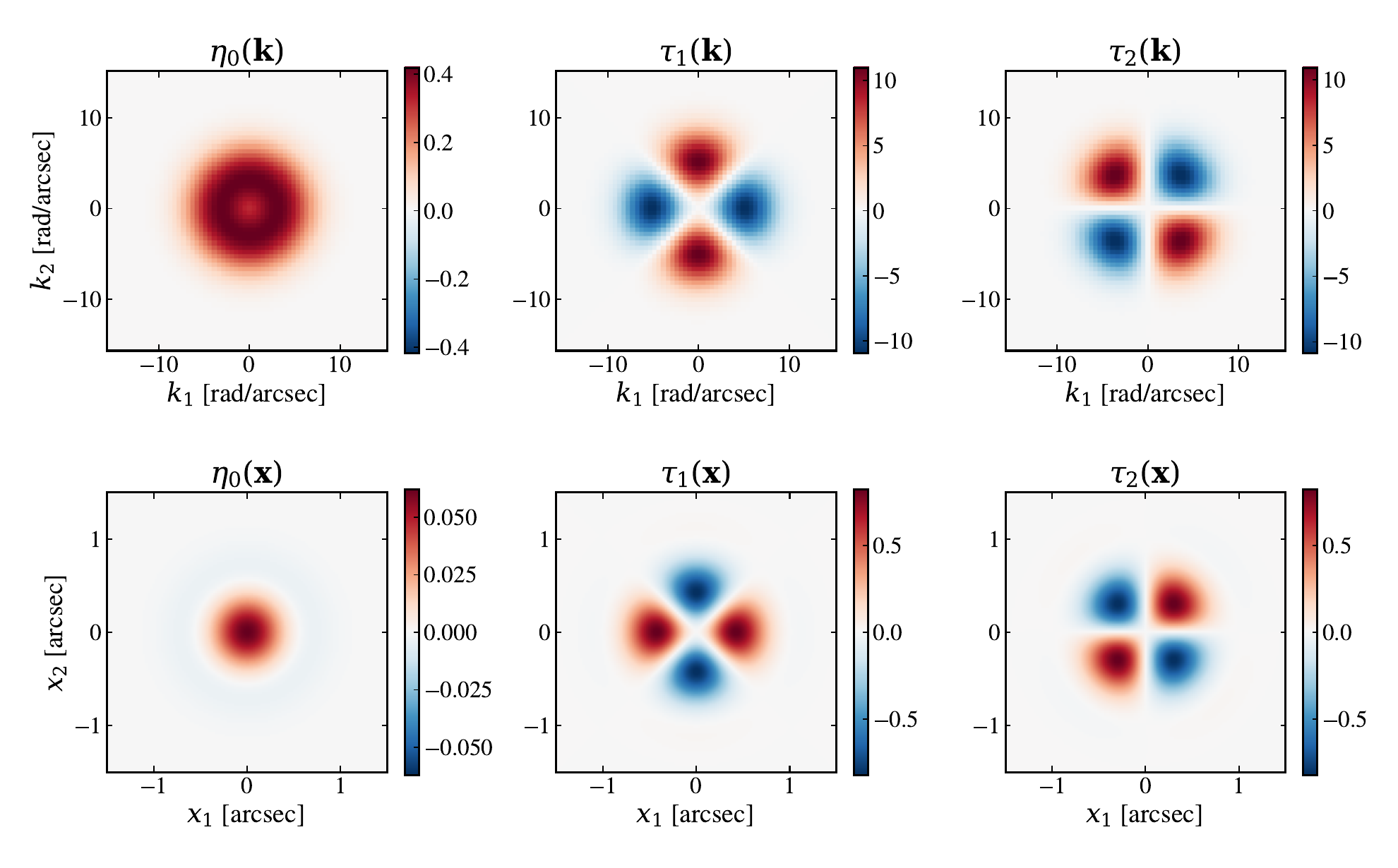}
  \caption{The added-noise kernel $\eta_0$ (equation~\ref{eq:nprime0}, left) and the two correction kernels $\tau_1$ (middle) and $\tau_2$ (right) of equation~(\ref{eq:tau}), in Fourier space (top) and in configuration space (bottom), for a Moffat PSF with $\beta=2.5$ and a full width at half maximum (FWHM) of $0.8''$ on $0.2''$ pixels, white image noise, $\sigma_h=0.4''$ and $\mu=0.1$. The configuration-space kernels are shown within $1.5''$ of the centre, interpolated to a grid eight times finer than the pixels by zero-padding in Fourier space; their values are per $0.2''$ pixel. $\eta_0$ depends on $|\vk|$ only, while $\tau_1$ and $\tau_2$ carry the quadrupole structure of $a_1$ and $a_2$ in equations~(\ref{eq:adef}) and~(\ref{eq:a2def}), the two differing by a $45^\circ$ rotation.}
  \label{fig:kernels2d}
\end{figure*}

We first make explicit what the pixel shear-response operator $\Kop$ does to the spectrum of the noise, and hence what the added noise has to undo. For Gaussian noise the distribution of $\zeta_\gamma$ is fixed by its covariance, so the condition of Section~\ref{sec:condition} becomes a statement about power spectra. Transforming the noise with $\Kop$ changes its power at every $\vk$ by the fraction $\gamma_1 a_1(\vk)$ (equation~\ref{eq:Sgamma}), so, writing $\Nadd_\gamma(\vk)$ for the power spectrum of $n^h_\gamma$, the condition reads
\begin{align}
  \frac{\partial}{\partial\gamma_1}\Big[\Snoise_\gamma(\vk)+\Nadd_\gamma(\vk)\Big]_{\gamma=0}&=0
  \nonumber\\
  \Longleftrightarrow\quad
  \left.\frac{\partial \Nadd_\gamma}{\partial\gamma_1}\right|_{0} &= -\Snoise(\vk)\,a_1(\vk),
  \label{eq:gausscond}
\end{align}
at every $\vk$ within the band of the smoothed image: the added spectrum must change in the direction opposite to that of the noise. A field with power $-\gamma_1\Snoise a_1$ alone cannot provide this, because a power spectrum must be non-negative for both signs of $\gamma_1$, which fails wherever $a_1(\vk)\neq0$. The condition fixes only the derivative of $\Nadd_\gamma$ at $\gamma=0$; its value there,
\begin{equation}
  \Nadd_0(\vk) > 0 \text{ wherever } a_1(\vk)\neq0,
  \label{eq:family}
\end{equation}
is free apart from this positivity, which is needed so that a non-negative family with the required derivative exists around $\gamma=0$. The value $\Nadd_0$ is the price of the cancellation: it is the noise that is actually added to the data, since the measurement is made at $\gamma=0$. Its amplitude, written below as the fraction $\mu$ of the original noise power, is not tied to the lensing shear, and we adopt the small value $\mu=0.1$ (Section~\ref{sec:measured}). A non-negative family with the required derivative is obtained by writing the added spectrum as a square. The simplest choice, and the one we adopt, has a constant amplitude $\sqrt{\mu}$:
\begin{equation}
  \Nadd_\gamma(\vk) = \Snoise(\vk)\left[\sqrt{\mu}-\gamma_1\frac{a_1(\vk)}{2\sqrt{\mu}}\right]^2 ,
  \label{eq:Nopt2}
\end{equation}
manifestly non-negative, with derivative $-\Snoise a_1$ at $\gamma=0$ as equation~(\ref{eq:gausscond}) requires. At $\gamma=0$ it equals $\mu\,\Snoise$, the original spectrum scaled by $\mu$: the added noise is a scaled-down, independent copy of the original noise, and $\mu$ is the added variance as a fraction of the original noise variance, at every $\vk$.

As in rotation renoising, let $\epsilon'(\vx)$ be an independent realization of the image noise (Section~\ref{sec:review}), drawn from the noise model of the pipeline or from source-free regions of the survey. The convolution of $\epsilon'$ with a kernel $\eta$ is
\begin{equation}
  (\eta\ast \epsilon')(\vx) = \iint\dd^2x'\;\eta(\vx-\vx')\,\epsilon'(\vx'),
  \label{eq:conv}
\end{equation}
whose Fourier transform is $\eta(\vk)\,\epsilon'(\vk)$; here $\eta(\vx)$ is a fixed, deterministic filter on the pixel grid, and not a random field. The added noise in the smoothed image is
\begin{equation}
  n^h_\gamma(\vx) = (\eta_\gamma\ast \epsilon')(\vx),
  \label{eq:nprime}
\end{equation}
with the kernel chosen so that the spectrum $|\eta_\gamma|^2 S$ of $n^h_\gamma$ equals $\Nadd_\gamma$ of equation~(\ref{eq:Nopt2}),
\begin{equation}
  \eta_\gamma(\vk) = \frac{h(\vk)}{p(\vk)}\left[\sqrt{\mu}-\gamma_1\frac{a_1(\vk)}{2\sqrt{\mu}}\right],
  \label{eq:kernels}
\end{equation}
where $|h/p|^2S=\Snoise$ by equation~(\ref{eq:Sdec}). The kernel is linear in $\gamma_1$. The factor $h/p$ is the deconvolution by the PSF and the reconvolution with the Gaussian that every noise image undergoes in the pipeline (equation~\ref{eq:smoothednoise}), so $n^h_\gamma$ lives in the same plane as $f^h$ and enters the renoised image (\ref{eq:renoised}) directly. Equivalently, one may add to the observed image the pure noise image $n_\gamma$ with $h\,n_\gamma/p=n^h_\gamma$, in the same way as the rotated noise field $n=\Rop\epsilon'$ of rotation renoising, so the pipeline of \citet{li2025} is unchanged apart from the choice of $n$. The noise spectrum itself has cancelled from the kernel: the kernel depends on the noise only through $a_1(\vk)$, which is built from the logarithm of the deconvolved spectrum $\Sd$. The deconvolution $1/p$ and the derivative in $a_1$ grow at large $k$, but the Gaussian $h$, chosen broader than the PSF (Section~\ref{sec:observables}), suppresses them, so $\eta_\gamma$ stays bounded and the added noise is band-limited like $f^h$.

The second shear component is handled by the same construction. The operator $\Kopb_2$ changes the noise power by the fraction $\gamma_2 a_2(\vk)$ (equation~\ref{eq:a2def}), the analogue of equation~(\ref{eq:Sgamma}) in the frame rotated by $45^\circ$, so the condition of Section~\ref{sec:condition} for $\gamma_2$ is equation~(\ref{eq:gausscond}) with $a_1$ replaced by $a_2$. A single realization $\epsilon'$ can carry both components through a kernel linear in the shear,
\begin{equation}
  \eta_\gamma(\vk) = \frac{h(\vk)}{p(\vk)}\left[\sqrt{\mu}-\gamma_1\frac{a_1(\vk)}{2\sqrt{\mu}}-\gamma_2\frac{a_2(\vk)}{2\sqrt{\mu}}\right],
  \label{eq:kernels12}
\end{equation}
whose added-noise spectrum $|\eta_\gamma|^2 S=\Snoise\big[\sqrt{\mu}-\gamma_1 a_1/(2\sqrt \mu)-\gamma_2 a_2/(2\sqrt \mu)\big]^2$ has shear responses $-\Snoise a_1$ and $-\Snoise a_2$ at $\gamma=0$, as the conditions for the two components require, and still equals $\mu\,\Snoise$ at the measurement point $\gamma=0$.

Because $\Sd=S/|p|^2$ is the real, even power spectrum of a real field, $a_1$ and $a_2$ are real and even in $\vk$, and the factor $h/p$ is Hermitian, $(h/p)(-\vk)=(h/p)^{*}(\vk)$, for a real PSF. The kernel $\eta_\gamma(\vk)$ is therefore Hermitian, so its configuration-space filter $\eta_\gamma(\vx)$, and hence the added field $n^h_\gamma$, is real.

The noise actually added to the data is the $\gamma=0$ member of the family, with kernel $\eta_0\equiv\eta_{\gamma=0}=\sqrt{\mu}\,h/p$,
\begin{equation}
  n^h(\vk) = \eta_0(\vk)\,\epsilon'(\vk) = \sqrt{\mu}\,\frac{h(\vk)}{p(\vk)}\;\epsilon'(\vk),
  \label{eq:nprime0}
\end{equation}
with spectrum $\mu\,\Snoise$. This single field is what the measurement adds to each image, and the shear responses of the next section are constructed entirely from it.

\subsection{Shear response of shapes from renoised images}
\label{sec:measured}

\begin{figure*}
  \centering
  \includegraphics[width=0.82\textwidth]{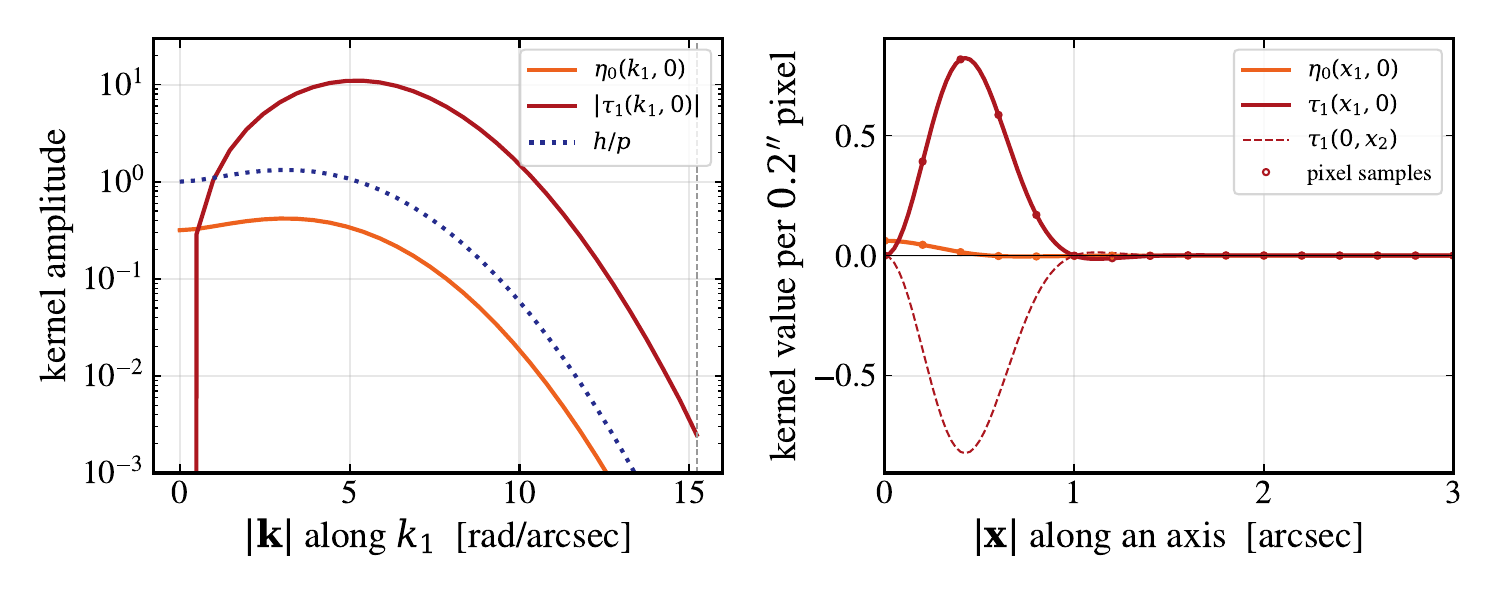}
  \caption{Profiles of the kernels of Figure~\ref{fig:kernels2d}. Left: $\eta_0$ and $|\tau_1|$ along the $k_1$ axis, where $\tau_1$ is negative, together with the deconvolution and smoothing factor $h/p$ that turns the image noise into the reconvolved noise $\epsilon^h$; the dashed vertical line marks the band limit $k_{\max}$ that \anacal{} applies in its deconvolution, which is not applied here. Right: $\eta_0$ and $\tau_1$ along the $x_1$ axis, and $\tau_1$ along the $x_2$ axis, interpolated to a grid eight times finer than the pixels by zero-padding in Fourier space; circles mark the values on the pixels.}
  \label{fig:kernels1d}
\end{figure*}

The renoised response estimator of equation~(\ref{eq:main}) needs, for each shear component, the pixel shear response of the added noise and its assigned response, both built from the single field $n^h$ of equation~(\ref{eq:nprime0}). The pixel shear responses $\Kop n^h$ and $\Kopb_2 n^h$ follow from equations~(\ref{eq:pixres})--(\ref{eq:pixres2}) with $n^h$ in place of $f^h$. The assigned responses follow by differentiating the kernel of equation~(\ref{eq:kernels12}) at $(\gamma_1, \gamma_2)=0$,
\begin{align}
  \dot n^h_{(1)}(\vk) &= -\frac{h(\vk)}{p(\vk)}\,\frac{a_1(\vk)}{2\sqrt{\mu}}\;\epsilon'(\vk) = -\frac{a_1(\vk)}{2\mu}\,n^h(\vk),
  \nonumber\\
  \dot n^h_{(2)}(\vk) &= -\frac{h(\vk)}{p(\vk)}\,\frac{a_2(\vk)}{2\sqrt{\mu}}\;\epsilon'(\vk) = -\frac{a_2(\vk)}{2\mu}\,n^h(\vk),
  \label{eq:assigned2}
\end{align}
so that the family is $n^h_{\gamma}=n^h+\gamma_1\dot n^h_{(1)}+\gamma_2\dot n^h_{(2)}$. Unlike rotation renoising, the assigned and pixel responses are not proportional: the assigned response is defined by the fixed spectral filter of equation~(\ref{eq:assigned2}), not by an opposite shear of the added field.

Following the response of equation~(\ref{eq:main}), the added noise enters the two shear responses only through the correction fields $\delta n^h_{(1)}\equiv\dot n^h_{(1)}-\Kop n^h$ and $\delta n^h_{(2)}\equiv\dot n^h_{(2)}-\Kopb_2 n^h$,
\begin{align}
  \avg{R_1} &= \avg{\sum_{\vx} e_{1,\vx}\big(\Kop f^h+\delta n^h_{(1)}\big)(\vx)},
  \label{eq:corr}\\
  \avg{R_2} &= \avg{\sum_{\vx} e_{2,\vx}\big(\Kopb_2 f^h+\delta n^h_{(2)}\big)(\vx)},
  \label{eq:main2}
\end{align}
and the shear estimators are $\hat\gamma_1=\avg{e_1}/\avg{R_1}$ and $\hat\gamma_2=\avg{e_2}/\avg{R_2}$. In configuration space each correction field is a homogeneous filter of $\epsilon'$ plus the coordinate terms of the pixel response,
\begin{align}
  \delta n^h_{(1)}(\vx) &= (\tau_1\ast\epsilon')(\vx) + x_1\frac{\partial n^h}{\partial x_1} - x_2\frac{\partial n^h}{\partial x_2},
  \label{eq:corrfield}\\
  \delta n^h_{(2)}(\vx) &= (\tau_2\ast\epsilon')(\vx) + x_1\frac{\partial n^h}{\partial x_2} + x_2\frac{\partial n^h}{\partial x_1},
  \nonumber
\end{align}
with the homogeneous kernels
\begin{align}
  \tau_1(\vk) &= -\frac{h(\vk)}{p(\vk)}\left[\frac{a_1(\vk)}{2\sqrt{\mu}}+\sigma_h^2\big(k_1^2-k_2^2\big)\sqrt{\mu}\right],
  \label{eq:tau}\\
  \tau_2(\vk) &= -\frac{h(\vk)}{p(\vk)}\left[\frac{a_2(\vk)}{2\sqrt{\mu}}+2\sigma_h^2k_1k_2\sqrt{\mu}\right].
  \nonumber
\end{align}
Each kernel combines the assigned response with the Gaussian term of the pixel response; the coordinate terms of equation~(\ref{eq:corrfield}), the $x_j\,\partial n^h/\partial x_k$ terms that carry an explicit factor of the position $\vx$, depend on the pixel's position relative to the centre about which the pixel shear response is evaluated and cannot be absorbed into a kernel, so they are computed from the stored $n^h$ exactly as for $f^h$. The kernels $\eta_0$, $\tau_1$ and $\tau_2$ are shown in Figure~\ref{fig:kernels2d}, with the axial profiles of $\eta_0$ and $\tau_1$ in Figure~\ref{fig:kernels1d}, for a representative ground-based configuration: a Moffat PSF \citep{moffat1969} with $\beta=2.5$ and a full width at half maximum of $0.8''$, sampled on $0.2''$ pixels, white image noise, and a smoothing Gaussian of $\sigma_h=0.4''$. This smoothing scale, $\sigma_h=0.4''$, and the added-noise fraction $\mu=0.1$ are the default values adopted throughout this work. In Fourier space $\eta_0$ is $\sqrt{\mu}$ times the deconvolution and smoothing factor $h/p$ of the pipeline: it depends on $|\vk|$ only, equals $\sqrt{\mu}$ at $\vk=0$, rises by about $30$ per cent where $1/p$ grows faster than $h$ decays, and is then cut off by the Gaussian; $\tau_1$ is the quadrupole pattern inherited from $a_1$, negative along $k_1$ and positive along $k_2$, peaking near $|\vk|\simeq5.4\,{\rm rad\,arcsec^{-1}}$. In configuration space both kernels are compact: $\eta_0$ is $\sqrt{\mu}$ times the kernel that turns the image noise into the reconvolved noise, a positive core about $0.3''$ wide with a weak negative ring, and $\tau_1$ has four lobes of alternating sign peaking at $|\vx|\simeq0.4''$, close to the smoothing scale. The second correction kernel $\tau_2$ is this same pattern rotated by $45^\circ$, inherited from $a_2$ of equation~(\ref{eq:a2def}), with lobes along the diagonals and the same peak amplitude. The kernels do not depend on the noise level, only on the PSF and on $\sigma_h$, because $S$ cancels in $a_1$ and $a_2$ for white noise.

\begin{figure*}
  \centering
  \includegraphics[width=0.49\textwidth]{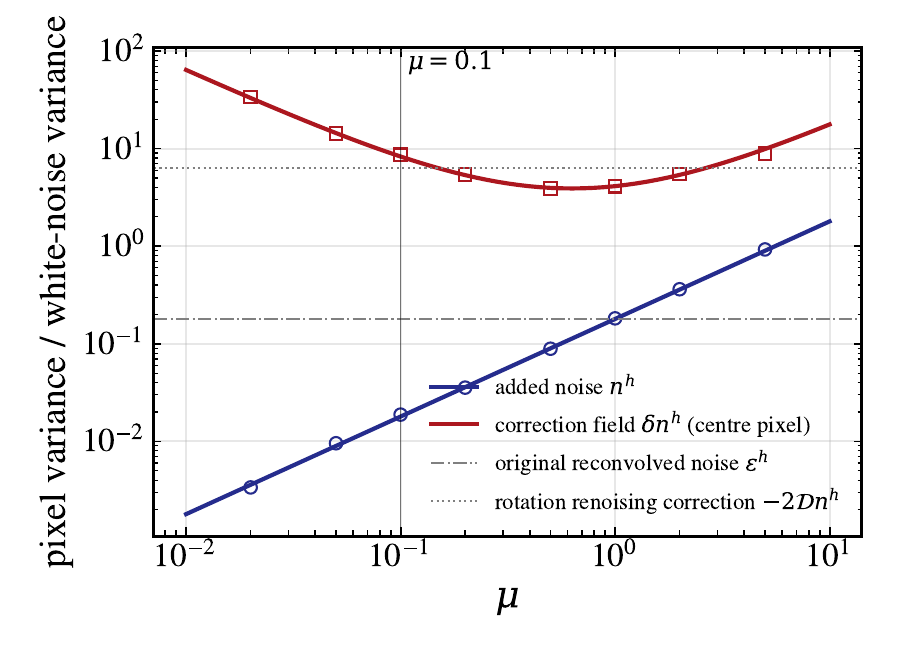}\hfill
  \includegraphics[width=0.49\textwidth]{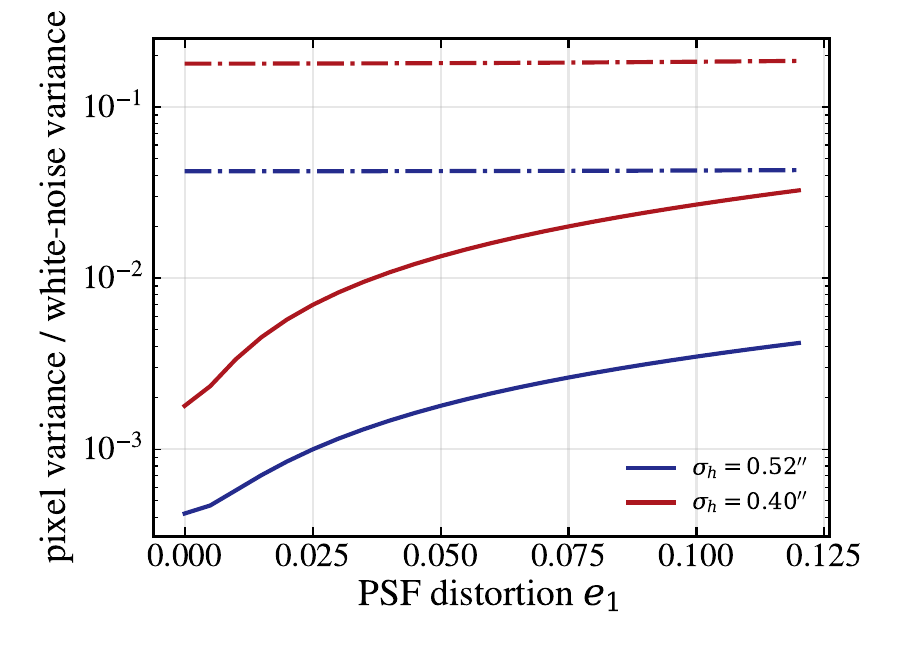}
  \caption{Pixel variances of the renoising fields, in units of the variance of the white image noise, for a Moffat PSF ($\beta=2.5$, $\mathrm{FWHM}=0.8''$) on $0.2''$ pixels. \emph{Left:} the added noise $n^h$ (equation~\ref{eq:nprime0}, blue) and the correction field $\delta n^h_{(1)}$ (equation~\ref{eq:corrfield}, red) at the galaxy centre, where its coordinate terms vanish, as a function of $\mu$, for the configuration of Figure~\ref{fig:kernels2d}. The vertical line marks the adopted value $\mu=0.1$; symbols are Monte Carlo estimates; the horizontal lines give the variance of the reconvolved original noise $\epsilon^h$, which is also the added noise of rotation renoising, and of the rotation-renoising correction $-2\Kop n^h$ at the galaxy centre. \emph{Right:} the variance of $\varphi^h$, the independent noise field added to cancel the spin-2 anisotropy that an anisotropic PSF imprints on the reconvolved noise (solid; Section~\ref{sec:spin2}), as a function of the PSF distortion $e_1$, for the two smoothing scales $\sigma_h$ and $\delta=0.01$; the dash-dot lines give the variance of the reconvolved noise $\epsilon^h$ for the same $\sigma_h$. The added field stays well below $\epsilon^h$, by a factor set by the PSF anisotropy.}
  \label{fig:variance}
\end{figure*}

For a spectrum invariant under rotations, as in the configuration of Figure~\ref{fig:kernels2d}, $a_2$ is $a_1$ rotated by $45^\circ$ and the two components carry the same noise cost; for a general anisotropic spectrum the two differ at the level of the anisotropy. On every quantity the added noise is the fraction $\mu$ of the original noise, against unity for rotation renoising on a $90^\circ$-symmetric spectrum (to which Section~\ref{sec:spin2} reduces the general case), so that only $10$ per cent of the original noise variance is added at the default $\mu$ (Figure~\ref{fig:variance}).

\subsection{Removing the spin-2 part of the unsheared noise spectrum}
\label{sec:spin2}

The shear estimator is accurate to second order only if the total noise at $\gamma=0$ is invariant under $90^\circ$ rotations, by the spin argument of LM24 (their eqs.~16--18); equivalently, the measured spin-2 ellipticity must carry no additive bias correlated with the spin-2 anisotropy of the noise. With the constant amplitude $\sqrt{\mu}$ of equation~(\ref{eq:Nopt2}) the total spectrum at $\gamma=0$ is $(1+\mu)\Snoise$, symmetric only if $\Snoise$ is; in general an anisotropic PSF makes $\Snoise=|h|^2\Sd$ differ between a mode $\vk$ and the rotated mode $\Rop\vk$, even for white image noise. We call $\Snoise(\vk)-\Snoise(\Rop\vk)$ the spin-2 part of the spectrum, for brevity: it is the part odd under $90^\circ$ rotation, which carries the angular harmonics of spin $2, 6, 10, \ldots$ ($m\equiv2\bmod4$); it changes sign under $\vk\to\Rop\vk$, since $\Rop^2\vk=-\vk$ and the power spectrum of a real field is even, $\Snoise(-\vk)=\Snoise(\vk)$. We remove it with little added noise and then run the construction of Sections~\ref{sec:construction}--\ref{sec:measured} on the symmetrized noise.

The smallest symmetric spectrum an independent filler can reach is $\max[\Snoise(\vk),\Snoise(\Rop\vk)]$ at every $\vk$, lifting the weaker of the two directions to the stronger. Because the spectrum is differentiated in the shear response calculation, we use a smooth maximum, defined in the deconvolved plane where $h$ cancels,
\begin{align}
  \Ssym(\vk) &= |h(\vk)|^2\,\Sd_{\rm sym}(\vk),
  \nonumber\\
  \Sd_{\rm sym} &= \frac12\Big[\Sd+\Sd_{\rm R}+\sqrt{\big(\Sd-\Sd_{\rm R}\big)^2+\delta^2\big(\Sd+\Sd_{\rm R}\big)^2}\,\Big],
  \label{eq:fill}
\end{align}
with $\Sd_{\rm R}(\vk)\equiv\Sd(\Rop\vk)$ and a small dimensionless $\delta$, for which we adopt $\delta=0.01$: it is exactly invariant under $\vk\to\Rop\vk$ and tends to the hard maximum $\max(\Sd,\Sd_{\rm R})$ as $\delta\to0$. The filler is built from a second pure-noise realization $\epsilon''$---drawn like $\epsilon'$, with the same correlation as the image noise, but independent of both $\epsilon$ and $\epsilon'$---through a fixed filter $\rho$, applied in Fourier space together with the deconvolution and re-smoothing that every noise image undergoes:
\begin{align}
  \varphi^h(\vk)&=\rho(\vk)\,\frac{h(\vk)}{p(\vk)}\,\epsilon''(\vk),
  \qquad
  \rho(\vk)=\sqrt{\frac{\Sd_{\rm sym}(\vk)}{\Sd(\vk)}-1},
  \nonumber\\
  \avg{|\varphi^h(\vk)|^2}&=\Snoise(\vk)\,\rho^2(\vk)=\Ssym(\vk)-\Snoise(\vk).
  \label{eq:filler}
\end{align}
We add $\varphi^h$ to the smoothed image $f^h$, so that the image noise becomes $\epsilon^h_{\rm s}=\epsilon^h+\varphi^h$ with reconvolved spectrum $\Ssym$ (equivalently $\epsilon_{\rm s}=\epsilon+\varphi$ before smoothing, with $\varphi=\rho\ast\epsilon''$). Applying $\rho$ together with $h/p$ keeps the filler band-limited by $h$ and never forms the unsmoothed observed-plane field $\varphi$, whose spectrum $S\,\rho^2$ reaches into the tail of $1/|p|^2$; $\rho$ depends only on the PSF and the image-noise spectrum. Its cost relative to the variance of $\epsilon^h$ is of order the fractional PSF anisotropy, far below the unit cost of the full rotated field; the smoothing adds a floor of order $\delta$, and the cost rises once $\sigma_h$ reaches into the tail of $1/|p|^2$ where the two directions differ most, so $\sigma_h$ should stay well above the PSF scale (right panel of Figure~\ref{fig:variance}).

From here on the image noise is the homogeneous Gaussian field $\epsilon_{\rm s}=\epsilon+\varphi$ (and $\epsilon^h_{\rm s}=\epsilon^h+\varphi^h$), with reconvolved spectrum $\Ssym$, whose realization is as unknown as that of $\epsilon$ and on which the pixel shear-response operator acts as before. Everything in Sections~\ref{sec:general}--\ref{sec:measured} therefore applies verbatim with $\epsilon\to\epsilon_{\rm s}$, $\Snoise\to\Ssym$ and the kernels $a_1,a_2$ replaced by $a_{1,\rm sym},a_{2,\rm sym}$, the functions $a_1,a_2$ of equations~(\ref{eq:adef}) and~(\ref{eq:a2def}) evaluated on the symmetrized spectrum,
\begin{align}
  a_{1,\rm sym}(\vk)&\equiv\left(k_1\frac{\partial}{\partial k_1}-k_2\frac{\partial}{\partial k_2}\right)\ln\Sd_{\rm sym}(\vk),
  \nonumber\\
  a_{2,\rm sym}(\vk)&\equiv\left(k_1\frac{\partial}{\partial k_2}+k_2\frac{\partial}{\partial k_1}\right)\ln\Sd_{\rm sym}(\vk).
  \label{eq:asym}
\end{align}
They are computed by the chain rule of equation~(\ref{eq:asymchain}), exactly from Fourier transforms and with no finite differences. Since $\Sd_{\rm sym}$ has no spin-2 part, $a_{1,\rm sym}(\Rop\vk)=-a_{1,\rm sym}(\vk)$ and $a_{2,\rm sym}(\Rop\vk)=-a_{2,\rm sym}(\vk)$ exactly.

The added noise $n^h$ is drawn from $\epsilon'$ with the constant amplitude $\sqrt \mu$ relative to the symmetrized spectrum, so its spectrum is $\mu\,\Ssym$; the kernels of Sections~\ref{sec:construction}--\ref{sec:measured} are reshaped by the factor $\sqrt{\Sd_{\rm sym}/\Sd}$,
\begin{align}
  \eta_0&=\sqrt{\mu}\,\frac{h}{p}\sqrt{\frac{\Sd_{\rm sym}}{\Sd}},
  \nonumber\\
  \dot n^h_{(1)}&=-\frac{a_{1,\rm sym}}{2\mu}\,n^h,
  \qquad
  \dot n^h_{(2)}=-\frac{a_{2,\rm sym}}{2\mu}\,n^h,
  \nonumber\\
  \tau_1&=-\frac{h}{p}\sqrt{\frac{\Sd_{\rm sym}}{\Sd}}\left[\frac{a_{1,\rm sym}}{2\sqrt \mu}+\sigma_h^2\left(k_1^2-k_2^2\right)\sqrt \mu\right],
  \nonumber\\
  \tau_2&=-\frac{h}{p}\sqrt{\frac{\Sd_{\rm sym}}{\Sd}}\left[\frac{a_{2,\rm sym}}{2\sqrt \mu}+2\sigma_h^2k_1k_2\sqrt \mu\right].
  \label{eq:kernelssym}
\end{align}
The total spectrum is then $(1+\mu)\Ssym+\mathcal{O}(\gamma^2)$, independent of $\gamma$ and symmetric, for a total added variance (filler plus added noise) of $0.14$ of that of $\epsilon^h$ for $\mu=0.1$, a PSF distortion $0.02$ and $\delta=0.01$; when $\Snoise$ is already symmetric, $\Sd_{\rm sym}=\Sd$, $a_{1,\rm sym}=a_1$, $\varphi=0$, and the construction reduces to that of Section~\ref{sec:construction}. The filler $\varphi$ and the added noise $n^h$ must come from independent realizations, which is why the construction uses the two fields $\epsilon''$ and $\epsilon'$: if both were filtered from the same field, their cross-spectrum would add a shear-dependent term to the total spectrum and violate the condition.

\subsection{Implementation}
\label{sec:implementation}

On a pixel grid with discrete Fourier frequencies $(k_1,k_2)$, a known image-noise power spectrum $S$ (the transform of the known noise correlation function $C$), the PSF $p$ of the cell and the smoothing Gaussian $h$, the construction runs per cell---the fixed-size subregion of the image on which \anacal{} runs its detection and measurement:
\begin{enumerate}
  \item \emph{Remove the spin-2 part of the noise} (Section~\ref{sec:spin2}). Form the deconvolved spectrum $\Sd=S/|p|^2$ and its rotation $\Sd_{\rm R}=\Sd(\Rop\vk)$, obtained by rotating the grid about $\vk=0$; the symmetrized spectrum $\Sd_{\rm sym}$ of equation~(\ref{eq:fill}); and the filler filter $\rho$ of equation~(\ref{eq:filler}). Draw the second pure-noise realization $\epsilon''$, with the same correlation as the image noise and independent of $\epsilon'$, and form the smoothed filler as the Fourier filter $\varphi^h=\rho\,(h/p)\,\epsilon''$ applied to the cell's cutout of $\epsilon''$: one more multiplication on a transform of the kind the pipeline already performs, with the factor $h$ keeping the filler band-limited so that the unsmoothed deconvolution $1/p$ never acts alone. From here on the filler is part of the image noise: the smoothed noise is $\epsilon^h_{\rm s}=\epsilon^h+\varphi^h$ with spectrum $\Ssym$.
  \item \emph{Differentiate the symmetrized spectrum.} Compute $a_{1,\rm sym}$ and $a_{2,\rm sym}$ of equation~(\ref{eq:asym}) by the chain rule of equation~(\ref{eq:asymchain}) from the $k$-derivatives of $\Sd$ and $\Sd_{\rm R}$. These follow exactly from Fourier transforms, without finite differences: since $\partial_{k_j}S=-\iu\,\mathcal{F}[x_jC]$ and $\partial_{k_j}p=-\iu\,\mathcal{F}[x_jp]$, one has $\dot S_1=\Sd\,(a_{S,1}-a_{p,1})$ with $a_{S,1}=\mathrm{Im}\big[k_1\mathcal{F}[x_1C]-k_2\mathcal{F}[x_2C]\big]/S$ and $a_{p,1}=2\,\mathrm{Im}\big[(k_1\mathcal{F}[x_1p]-k_2\mathcal{F}[x_2p])/p\big]$; likewise $\dot S_2$ for the second component. Where the deconvolution floors $|p|$, $a_{p,1}=a_{p,2}=0$.
  \begin{figure*}
    \centering
    \includegraphics[width=\textwidth]{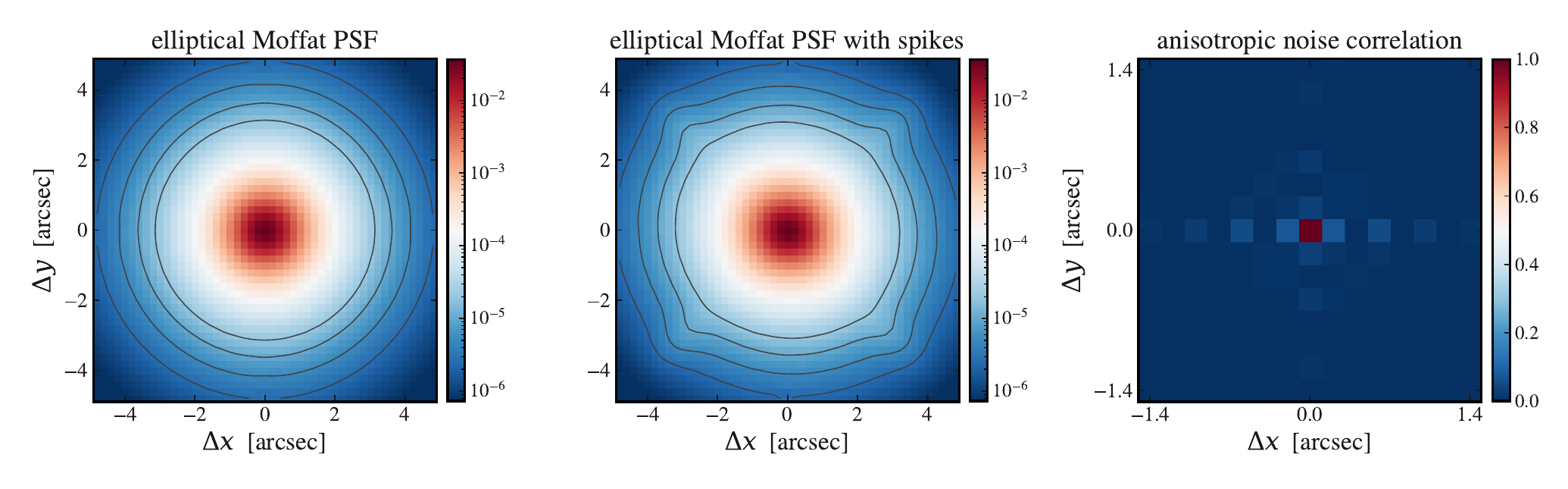}
    \caption{The PSF and noise inputs of the systematics tests, in arcsec at the $0.2''$ pixel scale. \emph{Left:} the elliptical Moffat PSF ($\mathrm{FWHM}=0.8''$, $\beta=2.5$) sheared to a distortion $e=(0.04,-0.04)$, on a logarithmic scale. \emph{Middle:} the same PSF with Rubin diffraction spikes added as a pure spin-4 pattern (four arms along $45^\circ/135^\circ$) carrying a fraction $0.032$ of the flux, which adds trace but no ellipticity; the two PSF panels share the colour scale and contour levels, and the spikes push the outer isophotes from ellipses towards a rounded square. \emph{Right:} the anisotropic noise correlation function, from stacking the measured autocorrelation of HSC coadd images and sheared to the same $e=(0.04,-0.04)$; the central $1.4''$ are shown on a linear scale.}
    \label{fig:psfnoise}
  \end{figure*}
  \item Form the kernels $\eta_0=\sqrt{\mu}\,h\,\sqrt{\Sd_{\rm sym}/S'}$, $\dot\eta_0=-\eta_0\,a_{1,\rm sym}/(2\mu)$ and $\dot\eta_0^{(2)}$ (the same with $a_{2,\rm sym}$) of equation~(\ref{eq:kernelssym}), where $S'$ is the spectrum from which the pure-noise image $\epsilon'$ was drawn ($S'=S$ for an unrotated draw, $S(\Rop\vk)$ for the rotated draw of the rotation-renoising pipeline). Multiply the transform of the cell's cutout of $\epsilon'$ by the three kernels to obtain $n^h$, $\dot n^h_{(1)}$ and $\dot n^h_{(2)}$. All kernels are Hermitian and suppressed by $h$ at large $k$, so the fields are real and band-limited. The variance of the smoothed noise in the cell is $(1+\mu)\intk|h|^2\Sd_{\rm sym}$, which replaces the factor two of rotation renoising in the detection thresholds and the measurement covariances.
  \item Form the first dimension (the pixel values) of the quintuple-number image of \citet{li2025}: add the filler $\varphi^h=\rho\,(h/p)\,\epsilon''$, together with $n^h=\eta_0\,\epsilon'$, to the smoothed image $f^h=h\,f^p/p$ of the data, as in equation~(\ref{eq:renoised})---three multiplications in Fourier space, on the transforms of the data and of the two stored noise realizations.
  \item Fill the remaining four dimensions with the response images. Since $\Kop$ and $\Kopb_2$ are linear, form the two shear-response images $\Kop f^h+\delta n^h_{(1)}=\Kop(f^h-n^h)+\dot n^h_{(1)}$ of equation~(\ref{eq:corr}) and $\Kopb_2 f^h+\delta n^h_{(2)}=\Kopb_2(f^h-n^h)+\dot n^h_{(2)}$ of equation~(\ref{eq:main2}) from equations~(\ref{eq:pixres})--(\ref{eq:pixres2}), in the same way as LM24 and \citet{li2025} form $\Kop(f^h-2n^h)$ and $\Kopb_2(f^h-2n^h)$; the last two dimensions hold the gradients $J_1$ and $J_2$ of equation~(\ref{eq:GJ}), computed from $f^h-n^h$, which carry the coordinate terms when the reference point of the response changes. Note that only the added noise $n^h$ is subtracted here: the filler $\varphi^h$ plays the role of image noise, whose realization is treated as unknown (Section~\ref{sec:spin2}), so it carries no assigned response and stays in $f^h$.
  \item Run detection and measurement on the quintuple-number image, as in LM24, Appendix~A: the shape of each detected galaxy, weighted by its detection weight, and the shear responses of the weighted shape follow automatically from the quintuple arithmetic. Evaluate equations~(\ref{eq:main}) and (\ref{eq:main2}), and $\hat\gamma_1=\avg{e_1}/\avg{R_1}$, $\hat\gamma_2=\avg{e_2}/\avg{R_2}$.
\end{enumerate}
In Fourier space the assigned responses are filters of the same transform of $\epsilon'$ that gives $n^h$, and they can be added to the Gaussian terms of $\Kop$ and $\Kopb_2$ before the inverse transforms that the pipeline already performs; none of the three kernels $\rho$, $\eta_0$, $\tau_j$ is ever realized in configuration space. The method therefore needs, beyond the transforms of rotation renoising, one transform per cell, of the cutout of $\epsilon''$, and the few transforms of $x_jC$ and $x_jp$ for the $k$-derivatives of the spectrum. A forced measurement of the detected sources on stamps uses the same steps with the stamp's PSF, the same image and the same cutouts of $\epsilon'$ and $\epsilon''$, and must use the same method as the detection; we have verified on simulated images that the cell-based and the stamp-based measurements of a source then agree to the numerical precision of the transforms for rotation and spectral renoising alike.

\section{Tests on image simulations}
\label{sec:tests}

\subsection{Simulation setup}
\label{sec:simsetup}

\begin{figure*}
  \centering
  \includegraphics[width=\textwidth]{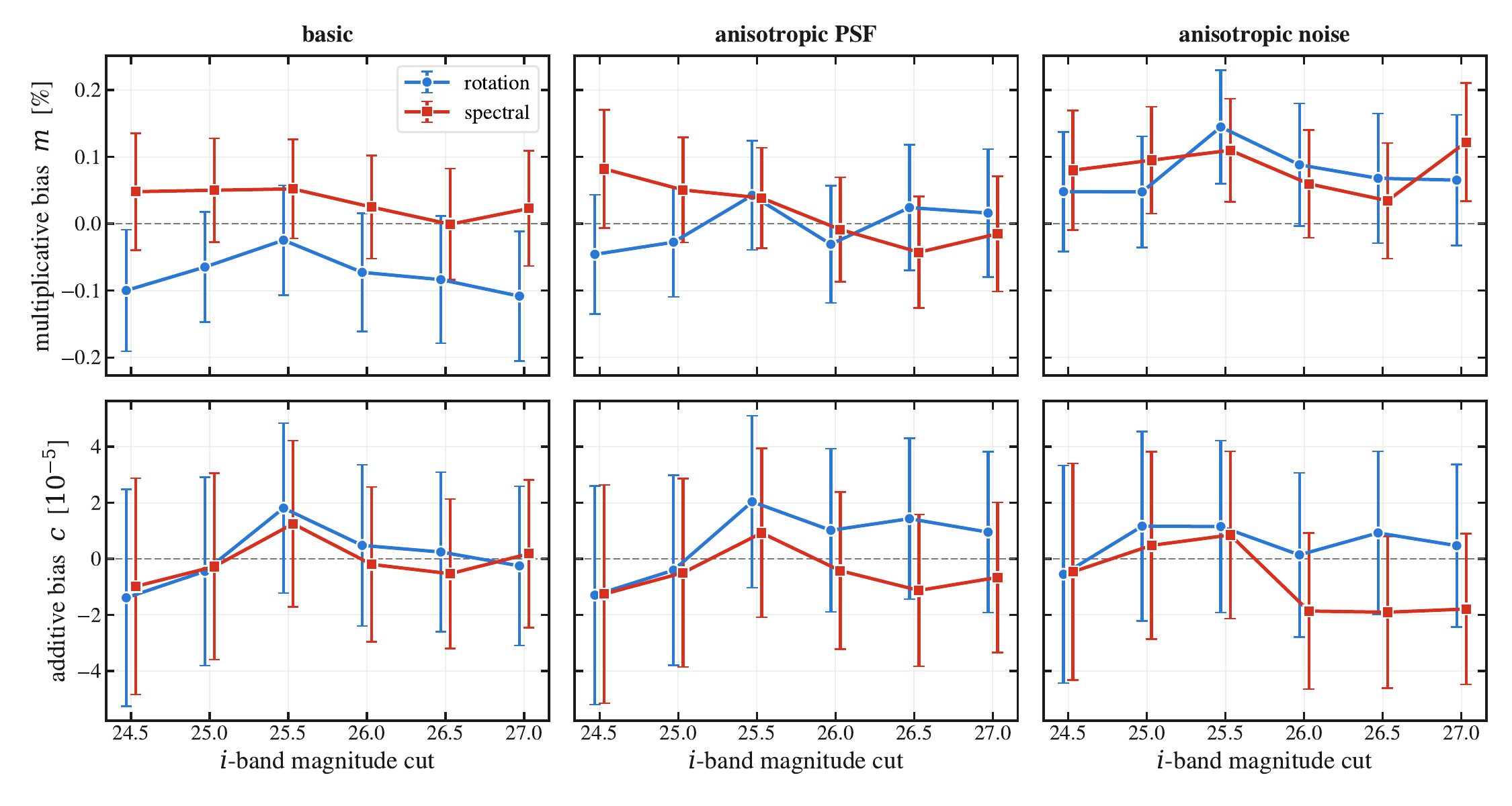}
  \caption{Multiplicative bias $m$ and additive bias $c_1$ of the recovered shear for rotation renoising (blue circles) and spectral renoising (orange squares), as a function of the $i$-band magnitude cut applied to the detected galaxies, for three test configurations of Section~\ref{sec:tests}: white noise with an isotropic PSF (the basic setup), an anisotropic PSF, and spatially correlated anisotropic noise. Error bars are the statistical uncertainty of the simulation and the dashed lines mark zero bias.}
  \label{fig:magcut_bias}
\end{figure*}

\begin{figure}
  \centering
  \includegraphics[width=\columnwidth]{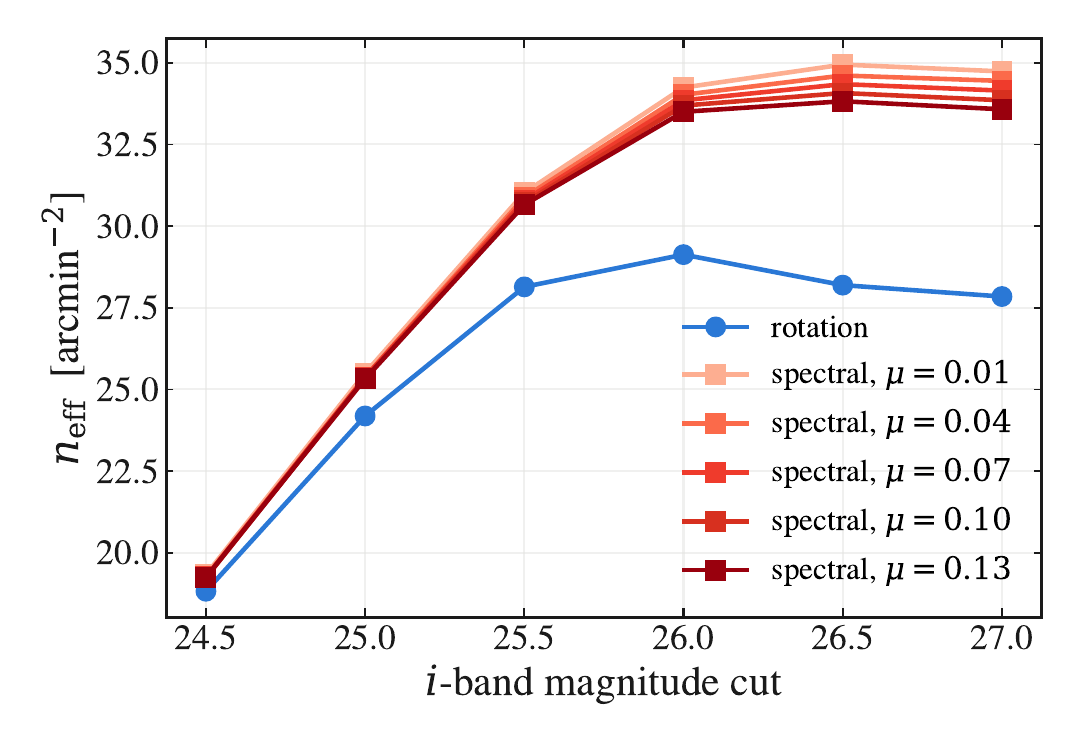}
  \caption{Effective galaxy number density $n_{\rm eff}$ of the galaxies passing the cut, as a function of the $i$-band magnitude cut, for rotation renoising (blue circles) and spectral renoising with the added-noise fraction $\mu$ from $0.01$ to $0.13$ (red squares, light to dark). The spectral curves lie within a few per cent of one another, so the gain in depth is insensitive to the choice of $\mu$.}
  \label{fig:magcut_neff}
\end{figure}

\begin{table}
\centering
\caption{Sky-background variance $\sigma^2_{{\rm sky},b}$ and the reciprocal effective Y10 coadd gain $t_b$ of equation~(\ref{eq:photonnoise}) for the three LSST bands of the simulation, at the photometric zero point of $30$.}
\label{tab:noise}
\begin{tabular}{lcc}
\hline
band & $\sigma^2_{{\rm sky},b}$ & $t_b$ \\
\hline
$r$ & $0.138$ & $1.04\times10^{-3}$ \\
$i$ & $0.354$ & $1.41\times10^{-3}$ \\
$z$ & $1.334$ & $2.31\times10^{-3}$ \\
\hline
\end{tabular}
\end{table}

\begin{table*}
\centering
\caption{Multiplicative bias $m$ and additive bias $c_1$ recovered for each
simulation case, for the two renoising methods. Each entry uses $10^{5}$
independent image pairs per shear mode, with the selection $i<26.0$, $|e|<1.5$ and
${\rm trace}>0.05$, where the trace is that of the measured second-moment
matrix. Uncertainties are $1\sigma$ seed bootstrap errors. All cases
are consistent with zero multiplicative bias at the $\lesssim2\sigma$ level.}
\label{tab:bias}
\begin{tabular}{lcccc}
\hline
case & $m$ (rotation) & $c_1$ (rotation) & $m$ (spectral) & $c_1$ (spectral) \\
     & $(10^{-3})$  & $(10^{-5})$  & $(10^{-3})$    & $(10^{-5})$ \\
\hline
basic                 & $-0.25\pm0.83$ & $1.81\pm3.03$ & $+0.52\pm0.74$ & $1.25\pm2.95$ \\
anisotropic PSF       & $+0.42\pm0.83$ & $2.04\pm3.07$ & $+0.39\pm0.76$ & $0.93\pm3.00$ \\
anisotropic noise     & $+1.45\pm0.85$ & $1.15\pm3.02$ & $+1.10\pm0.77$ & $0.85\pm2.96$ \\
diffraction-spike PSF & $+0.87\pm0.84$ & $1.94\pm3.08$ & $+0.76\pm0.77$ & $1.11\pm3.02$ \\
source photon noise   & $+0.14\pm0.83$ & $1.81\pm3.02$ & $+0.23\pm0.75$ & $1.31\pm2.95$ \\
\hline
\end{tabular}
\end{table*}

We test the two forms of renoising on the image simulations based on \citet{li2025}, with the FPFS shapelet shear estimator \citep{li2018,liLiMassey2022}, which propagates the pixel shear response~(\ref{eq:pixres}) through linear projections onto shapelet basis functions (Appendix~\ref{app:modes}). The images are generated with the \textsc{xlens} package \citep{sheldon2023,zhou2026}, built on \textsc{GalSim} \citep{rowe2015}, which produces background-subtracted, photometrically calibrated coadded images with their noise variance maps, world coordinate system and PSF model. The simulation covers the LSST $r$, $i$ and $z$ bands: galaxies are detected and their shapes measured on the $riz$ coadd of the three bands, while the magnitude cuts applied below are in the $i$ band. Sources of systematic error other than the noise bias, such as PSF modelling errors, noise misestimation, image warping and coaddition artefacts, are deliberately excluded, so that the test isolates the renoising step. \paragraph*{Input galaxies.} The input galaxies are bulge-plus-disc systems drawn from the Euclid Flagship galaxy mock catalogue \citep{euclidFlagship2025,potter2017}, whose joint distributions of magnitude, size and colour are calibrated to observations and extend well below the magnitude cuts applied here. Each galaxy is rendered as an exponential disc plus a S\'ersic bulge \citep{sersic1963}, with the bulge-to-total flux ratio, the component half-light radii, the bulge S\'ersic index and the disc inclination taken from the catalogue, whose morphologies are calibrated to the GOODS-South \citep{giavalisco2004} and CANDELS \citep{grogin2011,koekemoer2011} fields; the random disc inclinations give the sample a realistic spread of intrinsic ellipticities. Each galaxy carries a spectral energy distribution assigned by matching to COSMOS galaxies \citep{scoville2007} in colour, absolute magnitude and redshift, from which its $r$, $i$ and $z$ fluxes are drawn, so that the colours and their correlations with magnitude, size and morphology follow those of observed galaxies. They are placed at random positions without clustering, at a raw density of $260$ per arcmin$^2$ down to an $i$-band magnitude of $30$, and blends are included.

\paragraph*{PSF.} The fiducial PSF is a circular Moffat profile \citep{moffat1969},
\begin{equation}
  p(r) \propto \left[1+\left(\frac{r}{r_{\rm d}}\right)^{2}\right]^{-\beta},
  \qquad
  \mathrm{FWHM}=2\,r_{\rm d}\sqrt{2^{1/\beta}-1},
  \label{eq:moffat}
\end{equation}
with scale radius $r_{\rm d}$, $\beta=2.5$ and a full width at half maximum of $0.8''$. Two anisotropic variants are shown in Figure~\ref{fig:psfnoise}: an elliptical Moffat sheared to a distortion $e=(0.04,-0.04)$, and the same profile with Rubin-like diffraction spikes. The spikes form a pure spin-4 pattern of four arms along $45^\circ$ and $135^\circ$, matching the Rubin spider geometry of two parallel vane pairs. Each arm is the Fraunhofer diffraction pattern of a vane of width $w$, a single-slit profile as a function of the angular distance $\theta$ from the core along the arm,
\begin{equation}
  I(\theta) \propto \mathrm{sinc}^{2}\!\left(\frac{\theta}{\theta_{1}}\right),
  \qquad
  \mathrm{sinc}(x)=\frac{\sin\pi x}{\pi x},
  \qquad
  \theta_{1}=\frac{\lambda}{w},
  \label{eq:spike}
\end{equation}
with first zero at $\theta_{1}=\lambda/w\simeq2.9''$ for a vane width $w=50\,$mm at $\lambda=700\,$nm; the pattern is averaged over an $18\%$ bandpass and convolved with the Moffat. The spiked component carries a fraction $\alpha=0.032$ of the total flux, so it adds trace without ellipticity, pushing the outer isophotes from ellipses towards a rounded square.

\paragraph*{Noise.} The fiducial noise is stationary Gaussian sky-background noise at the Y10 depth \citep{lsstdesc2018}, at a photometric zero point of $30$. A correlated-noise test replaces the white sky noise by the anisotropic noise correlation function of Figure~\ref{fig:psfnoise} (right), formed by stacking the measured noise autocorrelation of HSC Public Data Release 3 \citep{aihara2022} coadd images \citep{li2022}, renormalized to unity at the origin, and sheared to $e=(0.04,-0.04)$; the same kernel drives both the injected noise and its bias correction. A photon-noise test adds the shot noise of the sources to the sky noise, so that the total variance of the pixel at $\vx$ is
\begin{equation}
  {\rm Var}[n(\vx)] = \sigma^2_{{\rm sky},b} + t_b\,f(\vx),
  \label{eq:photonnoise}
\end{equation}
with $f(\vx)$ the noiseless flux in the pixel, $b$ the photometric band, $\sigma^2_{{\rm sky},b}$ the sky-background variance in that band and $t_b$ the reciprocal of its effective Y10 coadd gain; their values at the simulation zero point of $30$ are listed in Table~\ref{tab:noise} for the three bands. The noise is then white but non-stationary: its variance tracks the galaxy light, reaching the sky level at a surface brightness of $i\simeq20.5\,{\rm mag\,arcsec^{-2}}$ and exceeding it only in the brightest cores, while reducing exactly to the sky-only field where there is no source flux. It is realized as a single standard-normal field scaled by $\sqrt{{\rm Var}[n(\vx)]}$, so that for a given seed the blank-sky pixels are bit-identical to the corresponding sky-only simulation and the two can be differenced pair by pair. This test probes the ground-based regime, where the sky background dominates and the source photon noise is a modest perturbation; the source-dominated noise of space-based surveys such as \textit{Euclid} \citep{laureijs2011} and Roman \citep{spergel2015} is left for future testing.

\paragraph*{Test protocol.} A uniform shear $\gamma_1=\pm0.02$ is applied to the profiles and positions of all galaxies in a subfield before convolution with the PSF and before the image noise is added; each scenario comprises $100\,000$ subfields of $21\,{\rm arcmin}^2$ each, $583\,{\rm deg}^2$ in total. To suppress shape noise, the shape-noise cancellation of \citet{pujol2019} is applied. Detection and selection are those of \citet{li2025}, with galaxy shapes measured by FPFS \citep{li2018,liLiMassey2022}, and magnitudes are those returned by the measurement; the only difference between the two runs is the renoising step. Rotation renoising adds $n=\Rop\epsilon'$ of Section~\ref{sec:rotation} and uses the response $\Kop(f^h-2n^h)$; spectral renoising adds $n^h=\eta_0\ast\epsilon'$ of equation~(\ref{eq:nprime0}) with $\mu=0.1$ and uses the response $\Kop(f^h-n^h)+\dot n^h_{(1)}$ of equation~(\ref{eq:corr}). The same realization $\epsilon'$ is used for both. We run the bias test in five configurations that probe the robustness of renoising to realistic departures from this fiducial setup: the \emph{basic} setup described above (white noise and an isotropic PSF); spatially correlated, anisotropic noise; an anisotropic PSF; a PSF with diffraction spikes; and photon shot noise from the sources at the Y10 depth. The anisotropic noise and the two anisotropic PSFs are shown in Figure~\ref{fig:psfnoise}.

\subsection{Results}
\label{sec:simresults}

From the shear estimated on the two sets of simulations, sheared by $\gamma_1=+0.02$ and by $\gamma_1=-0.02$ (Section~\ref{sec:simsetup}), we characterize the recovered shear by
\begin{equation}
  \hat\gamma_1 = (1+m)\,\gamma_1 + c_1 ,
  \label{eq:mc}
\end{equation}
which defines the multiplicative bias $m$ and the additive bias $c_1$ \citep{massey2007}. We measure them from the fields sheared by $\gamma_1=\pm0.02$ as
\begin{equation}
  m = \frac{\sum(e_1^{+}-e_1^{-})}{\sum(R_1^{+}+R_1^{-})}\,\frac{1}{\gamma_1}-1,
  \qquad
  c_1 = \frac{\sum(e_1^{+}+e_1^{-})}{\sum(R_1^{+}+R_1^{-})},
  \label{eq:mc_est}
\end{equation}
where $e_1^{\pm}$ and $R_1^{\pm}$ are the ellipticities and responses of the galaxies in the fields sheared by $\gamma_1=\pm0.02$ and the sums run over all galaxies passing the cut. We characterize the galaxy sample by the effective number density $n_{\rm eff}$, defined through the statistical error of the shear \citep{heymans2012, chang2013}: with $\hat\gamma_1^{(s)}$ the shear estimated from a single subfield of area $A$ and $\sigma_\gamma$ the dispersion of $\hat\gamma_1^{(s)}$ over the subfields,
\begin{equation}
  n_{\rm eff} = \frac{\sigma_{e}^2}{\sigma_\gamma^2\,A},
  \qquad \sigma_e=0.26,
  \label{eq:neff}
\end{equation}
the density of galaxies with intrinsic shape dispersion $\sigma_e$ per component and no measurement noise that would give the same shear error; $\sigma_e=0.26$ is the per-component intrinsic shape dispersion adopted for LSST forecasts \citep{chang2013, lsstdesc2018}. The dispersion $\sigma_\gamma$ includes the shape noise, the image noise and the added noise of each galaxy, so that $n_{\rm eff}$ is reduced by the measurement noise of the faint galaxies; we compute it from the fields of one shear sign, without the shape-noise cancellation, with $5\sigma$ clipping of outliers. Figure~\ref{fig:magcut_bias} shows the multiplicative bias $m$ and additive bias $c_1$ for the three configurations, and Figure~\ref{fig:magcut_neff} the effective number density $n_{\rm eff}$, as functions of the $i$-band magnitude cut from $24.5$ to $27$, for the two renoising schemes.

\begin{table}
\centering
\caption{Multiplicative bias $m$ and additive bias $c_1$ for spectral renoising as a function of the added-noise fraction $\mu$. Rotation renoising has no equivalent parameter, so only spectral renoising is shown. Each entry uses $10^{5}$ independent image pairs per shear mode, with the selection of Table~\ref{tab:bias}. Uncertainties are $1\sigma$ seed bootstrap errors. The $\mu=0.10$ row is the default used elsewhere in this work and coincides with the basic case of Table~\ref{tab:bias}. All values are consistent with zero multiplicative bias at the $\lesssim2\sigma$ level, and show no significant dependence on $\mu$ over the range tested.}
\label{tab:renoise_c}
\begin{tabular}{lcc}
\hline
$\mu$ & $m$ $(10^{-3})$ & $c_1$ $(10^{-5})$ \\
\hline
$0.01$           & $+1.28\pm0.80$ & $1.38\pm2.95$ \\
$0.04$           & $+1.50\pm0.75$ & $1.21\pm2.95$ \\
$0.07$           & $+0.61\pm0.74$ & $1.20\pm2.95$ \\
$0.10$ (default) & $+0.52\pm0.74$ & $1.25\pm2.95$ \\
$0.13$           & $+0.18\pm0.76$ & $1.32\pm2.94$ \\
\hline
\end{tabular}
\end{table}

The multiplicative bias is consistent with zero for both schemes at every cut and in all three configurations of Figure~\ref{fig:magcut_bias}, within a statistical uncertainty of $1$--$3\times10^{-3}$, and within the requirement $|m|<2\times10^{-3}$ of the LSST Dark Energy Science Collaboration for its ten-year analysis \citep{lsstdesc2018, sheldon2023}; the additive bias $c_1$ is likewise consistent with zero. Table~\ref{tab:bias} confirms this at higher statistics and for the two further cases: with $10^5$ image pairs per shear mode at the bright selection $i<26.0$, every case gives $|m|\lesssim1.5\times10^{-3}$, consistent with zero at the $\lesssim2\sigma$ level for both renoising schemes, while the additive bias $c_1$ stays at the $10^{-5}$ level and is consistent with zero. Spectral renoising shows no bias in excess of rotation renoising: its values of $m$ lie within $2\times10^{-3}$ of zero up to the faintest cut and within one standard deviation of the rotation-renoising values. Its accuracy is also insensitive to the added-noise fraction: Table~\ref{tab:renoise_c} gives $m$ and $c_1$ for $\mu$ from $0.01$ to $0.13$, all consistent with zero and with no significant dependence on $\mu$. This is the accuracy expected from Section~\ref{sec:condition}: both schemes satisfy the same first-order condition, so neither leaves a first-order noise response. Any residual is of second order in shear, or stems from departures from the assumed noise model---for spectral renoising, homogeneous Gaussian noise with known spectrum (Section~\ref{sec:spectral})---and both are below the precision of the simulation.

Figure~\ref{fig:magcut_neff} quantifies the depth gain. For the brightest cuts the two schemes give similar $n_{\rm eff}$ and similar raw densities $n_{\rm raw}$ of detected galaxies passing the cut: these galaxies are well above the noise and the added field hardly affects their measurement. The curves separate from $i\simeq25$ on. With rotation renoising $n_{\rm eff}$ peaks at $29$ per arcmin$^2$ at $i=26$ and declines to $28$ per arcmin$^2$ at $i=27$, because the doubled noise makes the faintest galaxies contribute less than they cost; with spectral renoising $n_{\rm eff}$ keeps rising, to $34$ per arcmin$^2$ at $i=26$ and $i=27$, a gain of about $20$ per cent at the faint end. The curves for $\mu$ between $0.01$ and $0.13$ lie within a few per cent of each other, so the gain is insensitive to the precise value of $\mu$. The raw density separates only at the faintest cuts, $110$ against $98$ per arcmin$^2$ at $i=27$, where the lower noise of the spectral scheme lets more faint galaxies be detected and pass the cut. The measured gain matches the expectation from the noise budget: LM24 attribute a loss of $20$--$25$ per cent in $n_{\rm eff}$ to the doubled noise of rotation renoising, and spectral renoising, which adds only $10$ per cent of the original variance, recovers it almost entirely.

\section{Summary and outlook}
\label{sec:discussion}

In this paper, we have derived the general condition under which an added noise field removes the noise bias of the shear response in the \anacal{} framework, and constructed a new solution of this condition, \emph{spectral renoising}, whose added variance is an adjustable small fraction of the original noise variance rather than a full extra copy. Our main results are as follows:

\begin{enumerate}
  \item We state the renoising condition (Section~\ref{sec:condition}): the distribution of the total noise---the original noise transformed with the pixel shear-response operator, plus the added noise carrying an \emph{assigned} shear dependence that we are free to choose---must be independent of shear to first order. Under this condition the shear response of the renoised ellipticity is expressed entirely in computable quantities (equation~\ref{eq:main}). The rotated noise field of \citet{sheldonHuff2017}, which we call rotation renoising, satisfies the condition and recovers the analytical response estimator of LM24 (Section~\ref{sec:rotation}).
  \item We construct spectral renoising (Section~\ref{sec:spectral}): for homogeneous Gaussian noise, the added field is an independent copy of the image noise scaled by $\sqrt{\mu}$, and its assigned shear response is the same field passed through a fixed filter built from the shear derivative of the logarithm of the PSF-deconvolved noise power spectrum. The added variance is the fraction $\mu$ of the original noise on every mode, against the full extra copy of rotation renoising; we adopt $\mu=0.1$. For an anisotropic PSF, the spin-2 part of the noise spectrum is first removed by a filler field, built from a second independent noise realization, whose cost is set by the PSF anisotropy (Section~\ref{sec:spin2}): the total added variance, filler plus added noise, is $0.14$ of that of the reconvolved noise for $\mu=0.1$, a PSF distortion of $0.02$ and $\delta=0.01$.
  \item We validate both renoising schemes on image simulations with realistic blending at the LSST Y10 depth, in five configurations: white noise with an isotropic PSF, spatially correlated anisotropic noise, an anisotropic PSF, a PSF with diffraction spikes, and source photon noise. With $10^5$ image pairs per shear mode, every case gives $|m|\lesssim1.5\times10^{-3}$, consistent with zero and within the requirement $|m|<2\times10^{-3}$ of the LSST Dark Energy Science Collaboration \citep{lsstdesc2018, sheldon2023}, while the additive bias stays at the $10^{-5}$ level (Figure~\ref{fig:magcut_bias}, Table~\ref{tab:bias}). The accuracy is insensitive to the added-noise fraction over $\mu=0.01$--$0.13$ (Table~\ref{tab:renoise_c}).
  \item Spectral renoising recovers the depth that rotation renoising loses to its doubled noise: the effective number density keeps rising with the magnitude cut, reaching $34$ per arcmin$^2$ at $i=27$ against $28$ for rotation renoising, a gain of about $20$ per cent at the faint end (Figure~\ref{fig:magcut_neff}), consistent with the $20$--$25$ per cent loss that LM24 attribute to the doubled noise.
\end{enumerate}

Spectral renoising is a drop-in replacement for the rotated noise field in the \anacal{} pipeline: the stored pure-noise image of the rotation-renoising pipeline can be reused unchanged, only the kernels applied to it change, at the cost of a few additional Fourier transforms per cell (Section~\ref{sec:implementation}).
\section*{Acknowledgements}

We thank Erin Sheldon for inspiring the idea of this work and for useful discussions.

This paper makes use of LSST Science Pipelines software developed by the Vera C.\ Rubin Observatory. We thank the Rubin Observatory for making their code available as free software at \url{https://pipelines.lsst.io}. This work also makes use of the open-source scientific Python ecosystem, including \texttt{NumPy} \citep{harris2020}, \texttt{SciPy} \citep{virtanen2020}, \texttt{Matplotlib} \citep{hunter2007}, \textsc{GalSim} \citep{rowe2015}, and the \texttt{conda-forge} distribution \citep{condaforge}.

Xiangchong Li acknowledges support from the U.S.\ Department of Energy (DOE) under Contract No.\ DE-SC0012704 and from the Laboratory Directed Research and Development Program at Brookhaven National Laboratory (Project No.\ 27992).

\section*{Data availability}

The \anacal{} code is publicly available on GitHub: \url{https://github.com/mr-superonion/AnaCal}. The image simulations are generated with the public \textsc{xlens} package \citep{zhou2026}, available at \url{https://github.com/mr-superonion/xlens}; the simulation outputs and the scripts producing the figures of this paper are available from the author on reasonable request.

\section*{AI usage disclosure}

Artificial intelligence (AI) tools have been used to assist with improving the English writing of this paper and with implementing the renoising method proposed here in the C++ code of \anacal{}. All scientific content, analysis, and conclusions are the author's own, and the author has reviewed and takes full responsibility for the final text and code.

\appendix

\section{Shear response of the noise spectrum}
\label{app:shear}

With the convention (\ref{eq:shear}), $\Mmat^{-\mathsf T}\vk=\big(k_1/(1-\gamma_1),\,k_2/(1+\gamma_1)\big)\simeq\big(k_1(1+\gamma_1),\,k_2(1-\gamma_1)\big)$, so that for any field $u$
\begin{equation}
  u\!\left(\Mmat^{-\mathsf T}\vk\right) = u(\vk) + \gamma_1\left(k_1\partial_{k_1}-k_2\partial_{k_2}\right)u(\vk)+\mathcal{O}(\gamma^2).
  \label{eq:appD}
\end{equation}
With the transform (\ref{eq:ft}), $\mathcal{F}^{-1}[k_j\partial_{k_j}u(\vk)]=-\partial_{x_j}\big(x_j u(\vx)\big)$; the terms $-u$ cancel between $j=1$ and $j=2$ and the remainder is $-x_1J_1+x_2J_2$, the coordinate terms of equation~(\ref{eq:pixres}). The smoothed image of the sheared galaxy is
\begin{align}
  f^h_\gamma(\vk) &= h(\vk)\,\frac{f^p}{p}\!\left(\Mmat^{-\mathsf T}\vk\right)
  = \frac{h(\vk)}{h(\Mmat^{-\mathsf T}\vk)}\,f^h\!\left(\Mmat^{-\mathsf T}\vk\right),
  \nonumber\\
  \frac{h(\vk)}{h(\Mmat^{-\mathsf T}\vk)} &= 1+\gamma_1\sigma_h^2\left(k_1^2-k_2^2\right)+\mathcal{O}(\gamma^2),
  \label{eq:appGauss}
\end{align}
because $|\Mmat^{-\mathsf T}\vk|^2=k^2+2\gamma_1(k_1^2-k_2^2)+\mathcal{O}(\gamma^2)$. The prefactor is the response of the Gaussian kernel to a shear applied between deconvolution and reconvolution, and gives the term $G_1$ of equation~(\ref{eq:pixres}); it is the Gaussian-kernel term of LM24, eq.~7. As a check of the relative sign of the two terms, take a round Gaussian galaxy of width $\sigma_g$, for which $f^h(\vk)=\exp[-k^2(\sigma_g^2+\sigma_h^2)/2]$. A shear of the galaxy alone changes its covariance to $\Mmat^{-1}\mathrm{diag}(\sigma_g^2)\Mmat^{-\mathsf T}$, and after the smoothing $f^h_\gamma(\vk)=f^h(\vk)\big[1-\gamma_1\sigma_g^2(k_1^2-k_2^2)\big]$: the stretch along $x_1$ narrows the transform along $k_1$. Equation~(\ref{eq:pixres}) gives $(k_1\partial_{k_1}-k_2\partial_{k_2})f^h=-(\sigma_g^2+\sigma_h^2)(k_1^2-k_2^2)f^h$ from the $J$ terms and $+\sigma_h^2(k_1^2-k_2^2)f^h$ from $G_1$, whose sum is the correct $-\sigma_g^2(k_1^2-k_2^2)f^h$ only with the opposite signs of the two terms as written; the Gaussian term must cancel the part of the coordinate term that shears the smoothing kernel. The same steps with $\gamma_1=0$ in equation~(\ref{eq:shear}), for which $\Mmat^{-\mathsf T}\vk\simeq(k_1+\gamma_2k_2,\,k_2+\gamma_2k_1)$, give the operator $(k_1\partial_{k_2}+k_2\partial_{k_1})$, the prefactor $1+2\gamma_2\sigma_h^2k_1k_2$ and hence $\Kopb_2$ of equation~(\ref{eq:pixres2}) and $a_2$ of equation~(\ref{eq:a2def}).

We now prove equation~(\ref{eq:Adef}) directly from the operator $\Kop$, using only that the noise is homogeneous. Let $\xi(\bm r)=\avg{\epsilon^h(\vx)\,\epsilon^h(\vx+\bm r)}$ be the correlation function of $\epsilon^h$, whose Fourier transform is $\Snoise(\vk)$, and split $\Kop=\Kopb_J+\Kopb_G$ into the coordinate part $\Kopb_J=-x_1\partial_{x_1}+x_2\partial_{x_2}$ and the Gaussian part $\Kopb_G$, the convolution whose transform is $\sigma_h^2(k_1^2-k_2^2)$. To first order the correlation function of $\epsilon^h+\gamma_1\Kop\epsilon^h$ between the points $\vx$ and $\vx'=\vx+\bm r$ is
\begin{equation}
  \xi(\bm r)+\gamma_1\Big[\avg{(\Kop\epsilon^h)(\vx)\,\epsilon^h(\vx')}+\avg{\epsilon^h(\vx)\,(\Kop\epsilon^h)(\vx')}\Big].
  \label{eq:appcorr}
\end{equation}
For the coordinate part, $\partial\xi(\vx'-\vx)/\partial x_j=-\partial\xi/\partial r_j$ and $\partial\xi(\vx'-\vx)/\partial x'_j=+\partial\xi/\partial r_j$, so the two cross terms are $(x_1\partial_{r_1}-x_2\partial_{r_2})\xi$ and $(-x'_1\partial_{r_1}+x'_2\partial_{r_2})\xi$, and their sum is
\begin{equation}
  -\left(r_1\frac{\partial}{\partial r_1}-r_2\frac{\partial}{\partial r_2}\right)\xi(\bm r).
  \label{eq:appJ}
\end{equation}
The explicit dependence on $\vx$ and $\vx'$ has cancelled: the transformed noise is still homogeneous, and the result does not depend on the point about which the shear is applied. With the transform (\ref{eq:ft}), $r_j\,\partial\xi/\partial r_j$ transforms to $-\partial(k_j\Snoise)/\partial k_j$, so expression~(\ref{eq:appJ}) transforms to $(k_1\partial_{k_1}-k_2\partial_{k_2})\Snoise$, the first term of equation~(\ref{eq:Adef}). For the Gaussian part, $(\Kopb_G\epsilon^h)(\vk)=\sigma_h^2(k_1^2-k_2^2)\,\epsilon^h(\vk)$ with a real factor, so the two cross terms in Fourier space give $2\sigma_h^2(k_1^2-k_2^2)\,\Snoise(\vk)$, the second term. Hence the spectrum of $\epsilon^h+\gamma_1\Kop\epsilon^h$ is
\begin{align}
  \Snoise_\gamma(\vk) &= \Snoise(\vk)
  +\gamma_1\Big[\left(k_1\partial_{k_1}-k_2\partial_{k_2}\right)\Snoise
  \nonumber\\
  &\quad+2\sigma_h^2\left(k_1^2-k_2^2\right)\Snoise\Big]+\mathcal{O}(\gamma^2),
\end{align}
which is equation~(\ref{eq:Adef}). Equation~(\ref{eq:adef}) follows by writing $\Snoise=|h|^2\Sd$, so that $\ln\Snoise=\ln\Sd-\sigma_h^2k^2$, and applying $(k_1\partial_{k_1}-k_2\partial_{k_2})$ to the logarithm: the derivative of $-\sigma_h^2k^2$ is $-2\sigma_h^2(k_1^2-k_2^2)$, which cancels the Gaussian term, leaving $a_1$.

\section{Shear derivative of the symmetrized spectrum}
\label{app:asym}

The symmetrized kernels $a_{1,\rm sym}$ and $a_{2,\rm sym}$ of equation~(\ref{eq:asym}) are the shear derivatives of $\ln\Sd_{\rm sym}$, the logarithm of the smooth maximum of equation~(\ref{eq:fill}). Because $\Sd_{\rm sym}$ is a nonlinear function of $\Sd$ and its rotation $\Sd_{\rm R}=\Sd(\Rop\vk)$, we differentiate it analytically by the chain rule, so that the $k$-derivatives act on the unsymmetrized spectra alone. Write
\begin{equation}
  \dot S_1\equiv\left(k_1\frac{\partial}{\partial k_1}-k_2\frac{\partial}{\partial k_2}\right)\Sd,
  \qquad
  \dot S_2\equiv\left(k_1\frac{\partial}{\partial k_2}+k_2\frac{\partial}{\partial k_1}\right)\Sd
  \label{eq:Sdot}
\end{equation}
for the shear derivatives of the deconvolved spectrum under the two spin-2 operators of equations~(\ref{eq:adef}) and~(\ref{eq:a2def}); the derivatives of the rotated spectrum are $\dot S_{1,\rm R}=-\dot S_1(\Rop\vk)$ and $\dot S_{2,\rm R}=-\dot S_2(\Rop\vk)$, with the minus sign because the operators are spin-2. With $d=\Sd-\Sd_{\rm R}$, $\sigma=\Sd+\Sd_{\rm R}$ and $D=\sqrt{d^2+\delta^2\sigma^2}$, the chain rule applied to equation~(\ref{eq:fill}) gives
\begin{align}
  a_{1,\rm sym}&=\frac{1}{2\Sd_{\rm sym}}\left[\dot S_1+\dot S_{1,\rm R}+\frac{d\,(\dot S_1-\dot S_{1,\rm R})+\delta^2\sigma\,(\dot S_1+\dot S_{1,\rm R})}{D}\right],
  \nonumber\\
  a_{2,\rm sym}&=\frac{1}{2\Sd_{\rm sym}}\left[\dot S_2+\dot S_{2,\rm R}+\frac{d\,(\dot S_2-\dot S_{2,\rm R})+\delta^2\sigma\,(\dot S_2+\dot S_{2,\rm R})}{D}\right].
  \label{eq:asymchain}
\end{align}
The derivatives $\dot S_1$, $\dot S_2$ and their rotations follow exactly from Fourier transforms, $\partial_{k_j}S=-\iu\,\mathcal{F}[x_jC]$ and $\partial_{k_j}p=-\iu\,\mathcal{F}[x_jp]$ (Section~\ref{sec:implementation}), so the symmetrized kernels are computed from the analytic derivative of the smooth maximum, with no numerical differencing.

\section{Linear modes and shapelets}
\label{app:modes}

The construction of Section~\ref{sec:spectral} does not depend on the Gaussian $h$. We summarize it here for the linear observables of the FPFS method \citep{li2018,liLiMassey2022} and LM24, the projections of the deconvolved image onto basis functions $\chi_i(\vk)$, such as polar shapelets or the detection kernels of LM24 (their eq.~1),
\begin{equation}
  \nu_i = \intk \frac{f^p(\vk)}{p(\vk)}\,\chi_i(\vk).
  \label{eq:appnu}
\end{equation}
The smoothed pixel values of Section~\ref{sec:observables} are the special case $\chi_{\vx}(\vk)=h(\vk)\,e^{\iu\vk\cdot\vx}$, for which $\nu_{\vx}=f^h(\vx)$ up to the normalization of the inverse transform. Under the shear (\ref{eq:shear}) the deconvolved image transforms as $(f^p/p)(\vk)\to(f^p/p)(\Mmat^{-\mathsf T}\vk)$, and a change of variables in equation~(\ref{eq:appnu}) moves the shear onto the basis function (LM24, eq.~6):
\begin{align}
  \chi_{;1i}(\vk) &= -\left(k_1\frac{\partial}{\partial k_1}-k_2\frac{\partial}{\partial k_2}\right)\chi_i(\vk),
  \nonumber\\
  \chi_{;2i}(\vk) &= -\left(k_1\frac{\partial}{\partial k_2}+k_2\frac{\partial}{\partial k_1}\right)\chi_i(\vk),
  \label{eq:appchiresp}
\end{align}
with the projections $\nu_{;1i}\equiv\intk(f^p/p)\,\chi_{;1i}$ and $\nu_{;2i}\equiv\intk(f^p/p)\,\chi_{;2i}$.
For the pixel basis, $\chi_{\vx;1}=[\sigma_h^2(k_1^2-k_2^2)-\iu(k_1x_1-k_2x_2)]\,h\,e^{\iu\vk\cdot\vx}$, which reproduces the Gaussian term $G_1$ and the coordinate terms of the pixel response $\Kop f^h$ of equation~(\ref{eq:pixres}). An ellipticity $e_1(\nv)$ is a nonlinear function of the modes, with response $R_1=\sum_i(\partial e_1/\partial\nu_i)\,\nu_{;1i}$.

The image noise $\epsilon$ contributes $\delta\nu_i=\intk(\epsilon/p)\chi_i$ to the modes, with covariance $\avg{\delta\nu_i\delta\nu_j^*}=\intk\Sd\chi_i\chi_j^*$, where $\Sd$ is the deconvolved noise spectrum of equation~(\ref{eq:Sdec}). The projection of the noise onto the response functions, $\delta\nu_{;1i}=\intk(\epsilon/p)\chi_{;1i}$, is the counterpart of $\Kop\epsilon^h$: the field $\epsilon/p$ would, if it were sheared with the galaxy, have the spectrum $\Sd(\Mmat^{-\mathsf T}\vk)=\Sd(1+\gamma_1a_1)$ with the same $a_1$ as in equation~(\ref{eq:adef}), and $\delta\nu_i+\gamma_1\delta\nu_{;1i}$ is its projection onto $\chi_i$. No Gaussian term appears, because there is no $h$ in the field. The renoised modes are $\tilde\nu_i=\nu_i+\delta\nu_i+\delta\nu'_i$ with $\delta\nu'_i=\intk(n/p)\chi_i$ for a pure noise image $n$ of known realization added to the observed image, and the response of the renoised ellipticity is (LM24, eq.~39)
\begin{equation}
  \avg{R_1} = \avg{\frac{\partial e_1}{\partial\tilde\nu_i}\left(\tilde\nu_{;1i}-\delta\nu_{;1i}-\delta\nu'_{;1i}\right)},
  \label{eq:appRrenoised}
\end{equation}
with $\tilde\nu_{;1i}$ the response measured on the renoised image; here and below, summation over the repeated mode index $i$ is implied. The condition of Section~\ref{sec:condition}, applied to the deconvolved field $(\epsilon/p)(\Mmat^{-\mathsf T}\vk)+n_\gamma/p$ for a family $n_\gamma$ with assigned response $\dot n_{(1)}=\partial n_\gamma/\partial\gamma_1|_0$, replaces the unmeasurable $-\delta\nu_{;1i}$ by $+\delta\nu'_{,\gamma i}\equiv\intk(\dot n_{(1)}/p)\chi_i$, which is the estimator (\ref{eq:main}) written on the modes. For Gaussian noise it requires the spectrum $\Nadd_\gamma$ of $n_\gamma/p$ to satisfy $\partial_{\gamma_1}\Nadd_\gamma|_0=-\Sd a_1$, equation~(\ref{eq:gausscond}) with $\Sd$ in place of $\Snoise$.

The squared family (\ref{eq:Nopt2}), with $\Sd$ in place of $\Snoise$, is $\Nadd_\gamma=\Sd(\sqrt \mu-\gamma_1 a_1/2\sqrt \mu)^2$, realized by filtering $\epsilon'$ with the kernels of equation~(\ref{eq:kernels}) without the factor $h/p$:
\begin{equation}
  n = \sqrt{\mu}\;\epsilon',
  \qquad
  \dot n_{(1)}(\vk) = -\frac{a_1(\vk)}{2\sqrt \mu}\,\epsilon'(\vk).
  \label{eq:appn}
\end{equation}
The added noise is $\sqrt \mu$ times $\epsilon'$, added to the observed image before deconvolution; the deconvolution and smoothing are supplied afterwards by $\chi_i/p$, as for the data, and $h\,n/p=n^h$ and $h\,\dot n_{(1)}/p=\dot n^h_{(1)}$ recover equations~(\ref{eq:nprime0}) and (\ref{eq:assigned2}). The two correction terms of the estimator combine into one kernel per mode,
\begin{align}
  \delta\nu'_{,\gamma i}-\delta\nu'_{;1i} &= \intk\frac{\epsilon'(\vk)}{p(\vk)}\,\tau_{1,i}(\vk),
  \nonumber\\
  \tau_{1,i}(\vk) &= -\left[\frac{a_1(\vk)}{2\sqrt \mu}\,\chi_i(\vk)+\sqrt \mu\,\chi_{;1i}(\vk)\right],
  \nonumber\\
  \tau_{2,i}(\vk) &= -\left[\frac{a_2(\vk)}{2\sqrt \mu}\,\chi_i(\vk)+\sqrt \mu\,\chi_{;2i}(\vk)\right],
  \label{eq:apptau}
\end{align}
the first term of each being the assigned response projected onto $\chi_i$ and the second the shear response of the basis function acting on the added noise. For the pixel basis, $\tau_{\vx}/p$ is the kernel $\tau_1$ of equation~(\ref{eq:tau}) together with the coordinate terms of equation~(\ref{eq:corrfield}). The estimator is then
\begin{equation}
  \avg{R_1} = \avg{\frac{\partial e_1}{\partial\tilde\nu_i}\left(\intk\frac{f^p+n}{p}\,\chi_{;1i}+\intk\frac{\epsilon'}{p}\,\tau_{1,i}\right)},
  \label{eq:appmain}
\end{equation}
where the first integral is the response $\tilde\nu_{;1i}$ measured on the renoised image and the second is a fixed linear functional of the stored realization $\epsilon'$; equivalently, moving the term $-\delta\nu'_{;1i}$ of $\tau_{1,i}$ into the first integral, the estimator is the response of the observed image alone---the renoised measurement minus the known response of $n$---plus the assigned response $\delta\nu'_{,\gamma i}$. The second shear component uses the same added noise $n$, with $\chi_{;2i}$, $a_2$ and $\tau_{2,i}$ in place of $\chi_{;1i}$, $a_1$ and $\tau_{1,i}$. When the spin-2 part of $\Sd$ is removed first (Section~\ref{sec:spin2}), the smoothed filler $\varphi^h=\rho\,(h/p)\,\epsilon''$, built from the second independent realization $\epsilon''$ (equation~\ref{eq:filler}), is added to $f^h$ and becomes part of the image noise, $\epsilon'$ is filtered by $\sqrt{\mu\,\Sd_{\rm sym}/\Sd}$ instead of $\sqrt \mu$ (by $\sqrt{\mu\,\Sd_{\rm sym}|p|^2/S'}$ if $\epsilon'$ was drawn from a spectrum $S'\neq S$), and $a_1$, $a_2$ are replaced by $a_{1,\rm sym}$, $a_{2,\rm sym}$ in the kernels $\tau_{1,i}$ and $\tau_{2,i}$; the Gaussian $h$ plays no role in any of these, so the modification is the same for every basis.

The costs follow mode by mode. The added variance is $\avg{|\delta\nu'_i|^2}=\mu\intk\Sd|\chi_i|^2$, the fraction $\mu$ of the original noise on every mode; the rotated noise field instead adds $\intk\Sd(\Rop\vk)\,|\chi_i|^2$, which equals the original mode variance when the deconvolved spectrum is $90^\circ$-symmetric. The correction variance is $\intk\Sd\,|a_1\chi_i/(2\sqrt \mu)+\sqrt \mu\,\chi_{;1i}|^2$. Minimizing it at each $\vk$ for a given basis function would replace the constant amplitude $\sqrt \mu$ by $\sqrt{a_1\chi_i/(2\chi_{;1i})}$ wherever this is real and positive; for the pixel basis at the galaxy centre, $\chi_{;1\vx}/\chi_{\vx}=\sigma_h^2(k_1^2-k_2^2)$. The Gaussian $h$ thus enters only through the choice of basis, and any band-limiting basis can be used with the same $n$, $\dot n_{(1)}$ and $a_1$.

\ifmnrasclass
  \bibliographystyle{mnras}
\else
  \bibliographystyle{aasjournal}
\fi
\bibliography{refs}

\end{document}